\documentclass[sigconf]{acmart}
\usepackage{multirow}
\usepackage{makecell}
\usepackage{listings}
\usepackage[most]{tcolorbox}
\usepackage[table]{xcolor}
\colorlet{mylightgray}{lightgray!70}

\AtBeginDocument{%
  }

\setcopyright{acmlicensed}
\copyrightyear{2018}
\acmYear{2018}
\acmDOI{XXXXXXX.XXXXXXX}
\acmConference[Conference acronym 'XX]{Make sure to enter the correct
  conference title from your rights confirmation email}{June 03--05,
  2018}{Woodstock, NY}
\acmISBN{978-1-4503-XXXX-X/2018/06}

\begin{document}

\title{AISA: Awakening Intrinsic Safety Awareness in Large Language Models against Jailbreak Attacks}

\author{Weiming Song}
\email{swm_academic@emails.bjut.edu.cn}
\affiliation{%
  \institution{Beijing University of Technology}
  \state{Beijing}
  \country{China}
}

\author{Xuan Xie}
\authornote{Corresponding authors.}
\email{xiexuan@must.edu.mo}
\affiliation{%
  \institution{Macau University of Science and Technology}
  \country{Macao}
}

\author{Ruiping Yin}
\authornotemark[1]
\email{yinruiping@bjut.edu.cn}
\affiliation{%
  \institution{Beijing University of Technology}
  \state{Beijing}
  \country{China}
}


\begin{abstract}
Large language models (LLMs) remain vulnerable to jailbreak prompts that elicit harmful or policy-violating outputs, while many existing defenses rely on expensive fine-tuning, intrusive prompt rewriting, or external guardrails that add latency and can degrade helpfulness. 
We present AISA, a lightweight, single-pass defense that activates safety behaviors already latent inside the model rather than treating safety as an add-on. 
AISA first localizes intrinsic safety awareness via spatiotemporal analysis and shows that intent-discriminative signals are broadly encoded, with especially strong separability appearing in the scaled dot-product outputs of specific attention heads near the final structural tokens before generation. 
Using a compact set of automatically selected heads, AISA extracts an interpretable prompt-risk score with minimal overhead, achieving detector-level performance competitive with strong proprietary baselines on small (7B) models. 
AISA then performs logits-level steering: it modulates the decoding distribution in proportion to the inferred risk, ranging from normal generation for benign prompts to calibrated refusal for high-risk requests—without changing model parameters, adding auxiliary modules, or requiring multi-pass inference. 
Extensive experiments spanning 13 datasets, 12 LLMs, and 14 baselines demonstrate that AISA improves robustness and transfer while preserving utility and reducing false refusals, enabling safer deployment even for weakly aligned or intentionally risky model variants.
\end{abstract}




\maketitle

\section{Introduction}
Large language models (LLMs) have achieved strong performance across many language understanding and generation tasks~\cite{matarazzo2025survey, huynh2025large, zhang2025exploring, he2025plan}.
A major barrier to their trustworthy deployment, however, is vulnerability to jailbreak attacks that bypass safety alignment and trigger harmful outputs~\cite{yi2024jailbreak, zou2023universal, chen2026jailbreaking, pu2025feint, liu2023autodan, dang2025rainbowplus}.
Existing defenses typically treat safety as an external constraint, enforced through resource-intensive fine-tuning~\cite{christiano2017deep, yuan2025refuse, liu2024adversarial}, invasive prompt modifications~\cite{jain2023baseline, zhang2025intention}, or static pre/post content filters~\cite{meta2024promptguard2, han2024wildguard}.
Some internal operations, such as activation steering~\cite{cao2025scans, wang2025steering}, also share this concept by imposing predefined safety directions on hidden representations, yet inevitably degrade model utility~\cite{wolf2025tradeoffs}.

The fundamental issue in this "safety‑as‑an‑add‑on" paradigm is that it contradicts the very purpose of building large models: after massive pre‑training, models should serve users directly~\cite{kojima2022large}.
When new modules and training techniques are continuously introduced to mitigate both the safety problems and the deficiencies in these modules themselves (e.g., alignment tax~\cite{lin2024mitigating}, over‑refusal~\cite{shi2024navigating}, inference latency~\cite{modelmetry2024latency}), user experience inevitably suffers~\cite{kim2024understanding}. 
This is also evidenced by widespread dissatisfaction in the
safety restrictions of recent commercial models like ChatGPT-5~\cite{ArsturnGPT5Debate}.

Consequently, we advocate for a shift toward minimal intervention.
We posit that an effective defense should satisfy five practical requirements: 
(1) operate within a single forward pass, 
(2) require no modifications to model parameters or user inputs, 
(3) impose negligible computational overhead, 
(4) rely on no large external components, and 
(5) preserve model utility on benign requests.
These requirements are especially relevant for platforms that cannot alter model parameters due to licensing or IP restrictions, but remain responsible for preventing harmful content generation~\cite{eu_ai_act_2024}.

In this paper, we propose \textbf{AISA}, a lightweight defense framework that meets all five requirements by awakening the intrinsic safety awareness in LLMs.
Our approach builds on the hypothesis that pre‑training instills a deep‑seated safety awareness that remains intact even when output generation seems compromised.
This intrinsic awareness, if properly accessed, could not only defend against various prompt-level jailbreak attacks but also enable the secure deployment of risky model variants.

We first show that even uncensored models~\cite{Hartford2025Uncensored} (fine‑tuned without safety constraints to maximize helpfulness) display refusal behavior on malicious requests.
Then, attempt to localize this awareness within the LLM's architecture.
Through a spatiotemporal search, we find that the ability to distinguish prompt intent is not confined to isolated components but is widely encoded throughout almost all internal activation spaces (spanning self-attention modules, feedforward networks, and residual connections).
Further analysis reveals that the optimal extraction point lies in the scaled dot‑product output space of attention heads, and the resulting prediction exhibits temporal convergence at structural tokens (e.g., control tokens) just before generation, providing a timely and stable safety signal.

AISA strengthens the discriminative capacity of its safety signal using a compact synthetic training set and automatically identifies the most informative attention heads through a simple data-driven ranking procedure, eliminating the need for costly or heuristic search. 
With a sparse selection of only 16 heads (0.004 MB in 16-bit precision) and constant O(1) runtime overhead that does not grow with model size, AISA extracts a robust safety signal that matches the detection performance of GPT‑5‑mini on 7B models.

Because this signal provides a real-time, numeric, and interpretable estimate of prompt intent, we use it to steer decoding directly by adjusting the logits toward safe behavior at the onset of generation. 
This intervention preserves semantic coherence: for harmful prompts, the base model already assigns elevated probability mass to refusal- or safety-related tokens, but standard decoding may fail to select them when these tokens remain sub-dominant. 
Our logits steering simply amplifies an existing preference rather than imposing a new behavior. 
Moreover, the continuous nature of the signal enables fine-grained control, allowing AISA to both reliably block high-severity harmful continuations—preventing “leakage” as tokens accumulate—and to provide calibrated, partial assistance when the harmful intent is ambiguous or lower in severity.



We conduct extensive experiments across 13 datasets, 14 defense baselines, and 12 LLMs.
The dataset includes 9 algorithmic jailbreaks, automatically generated malicious prompts, adversarial benign/malicious samples, wild jailbreaks, and sensitive self‑harm prompts.
In terms of model coverage, we span the full alignment spectrum, including dark models (fine-tuned on forbidden materials for malicious purposes), uncensored models, standard instruction models, safety‑aligned checkpoints, and even variants where safety‑relevant activation directions are ablated.
We include both dense and modern Mixture-of-Experts (MoE) designs, with sizes ranging from 7B to 20B parameters across 4 model families (Llama, Mistral, Qwen, and GPT-oss).
We benchmark against nine detection models and seven defense approaches, and verify model performance on four standard utility tasks (MMLU, GSM8K, MMLUpro, and BOOLQ).
The results show that AISA outperforms existing baselines and gets better generalization, while preserving utility on benign queries.
Finally, we observe that the specialized dataset developed for AISA also enhances the efficacy of other methods, and non‑aligned models equipped with our method can match or even surpass their safety‑aligned counterparts.

Our contributions are summarized as follows:
\begin{itemize}
\item We propose AISA, a lightweight defense that awakens LLMs' intrinsic safety awareness. AISA extracts safety signals from attention heads via linear probing and steers generation towards safety through logits adjustment.
\item We provide an in-depth analysis of the intrinsic safety awareness of LLMs and distilled a set of implications towards utilizing in the defense.
\item We conduct extensive experiments across 13 datasets, 9 jailbreaks, and 12 LLMs, showing that AISA outperforms 14 baselines in accuracy and transferability, while also preserving utility on benign requests.
\end{itemize}

\section{Related Work}
\subsection{Jailbreak Attacks on LLMs}
In broad terms, any prompt crafted to induce an LLM to output content that violates safety policies can be characterized as a jailbreak attack~\cite{xu2024comprehensive, greshake2023not}.
Based on the prompt generation process, we categorize jailbreak attacks into three streams:
(1) \textit{In-the-wild Jailbreak Prompts}~\cite{shen2024anything, jiang2024wildteaming}, which are created by real users seeking to obtain malicious outputs from LLM platforms, whether motivated by curiosity or deliberate intent.
(2) \textit{Optimization-based Jailbreak Prompts}, which employ searching~\cite{dang2025rainbowplus, chao2025jailbreaking, liu2025autodan} or gradient-based methods~\cite{zou2023universal, zhou2025don} to refine a specific malicious prompt.
(3) \textit{Automated RedTeaming Prompts}, which deploys specialized agents for jailbreak generation, either by leveraging commercial LLMs~\cite{mazeika2024harmbench} or fine-tuning open-source models~\cite{lee2024learning, hong2024curiosity} dedicated to this task.

Our defense evaluation covers all three categories, with algorithmic jailbreaks alone spanning 9 variants.
While some white‑box attacks~\cite{arditi2024refusal, wang2024trojan} that modify model parameters fall outside the standard prompt‑based threat model and are rarely included in previous defense papers~\cite{cao2025scans, xie2023defending}, we show that AISA can still defend against them. 
However, our primary focus remains on prompt‑based attacks that reflect real‑world deployment risks.
More details regarding threat model assumptions are provided in Appendix~\ref{app:threat model}.

\subsection{Jailbreak Defense}
To counter the jailbreak attacks, various defense mechanisms have been developed.
Current mainstream approaches primarily fall into four categories.
(1)\textit{Detection-based methods}~\cite{candogan2025single, jackhhaoClassifier}:
These methods focus on identifying malicious intent from prompts or screening LLM-generated content.
However, malicious intent detectors often fail to generalize beyond their training distribution, while detecting output after generation incurs computational waste.
(2)\textit{Safety Alignment Fintuning}~\cite{qi2024safety, dai2023safe}:
This approach modifies model behavior through safety-oriented fine-tuning, typically employing techniques such as Reinforcement Learning from Human Feedback (RLHF) or direct preference optimization. While effective at building robust safety barriers, this method demands substantial computational resources and high-quality human feedback data.
(3)\textit{Prompt Modification}:
Compared to the previous approaches, this category offers a more lightweight solution. 
Representative strategies include: employing system prompts to constrain user input within safety boundaries~\cite{xie2023defending, xu2024safedecoding}, and rewriting or paraphrasing original prompts before feeding them to the model~\cite{ jain2023baseline}.
(4)\textit{Activation Steering}: These methods steer represenatations towards predefined safety directions~\cite{arditi2024refusal, sheng2025alphasteer}, but requires careful operations and often degrades model utility~\cite{wolf2025tradeoffs}.

Despite their merits, these approaches fail to satisfy all five requirements outlined in the introduction. 
Moreover, AISA uses detection signals to guide defense interventions, enabling dynamic responses to prompts of varying maliciousness levels.
This capability is rarely achieved in existing defenses.

\subsection{Mechanistic Analysis and LLM Safety}
Analyzing internal mechanisms to address model safety is a well-established direction~\cite{zou2023representation}.
Recent approaches mainly explore two strategies: (1) locating safety-related neurons within feed-forward networks~\cite{chen2025towards,weng2025safe}, and (2) injecting safety behaviors via residual connections~\cite{arditi2024refusal,sheng2025alphasteer}.
Our work differs in three aspects.
First, instead of in-place modification of internal activations, we extract safety signals to guide downstream decoding decisions.
Second, we identify attention heads as a more promising location based on empirical analysis, contrasting with existing works that focus elsewhere.
Third, while some works also examine attention heads, their approaches differ fundamentally.
SafetyHeadAttribution~\cite{zhou2025role} ablates heads to show their disruptive effect; we probe heads to demonstrate their defensive potential.
AHASI~\cite{arai2025jailbreak} still relies on a static internal intervention strategy; AISA, in contrast, exploits dynamic safety awareness to achieve superior generalizability against the diverse distributions of unseen prompts.

\section{Methodology}
\label{sec:section_3}
\subsection{Preliminaries}
Large language models are composed of successive transformer layers~\cite{vaswani2017attention}, with each layer iteratively refining the representation from its predecessor. 
At the core of each layer lie two fundamental components: the Multi-Head Self-Attention (\textit{self-attn}) module and the Feed-Forward Network (\textit{ffn}).
These components are connected via layer normalization and residual connections, forming a stable information pathway~\cite{matarazzo2025surveylargelanguagemodels}.
Architecturally, they are implemented as sequences of linear projections that shuttle representations across distinct, high-dimensional activation spaces.

Our method, AISA, builds on this architectural property by extracting a safety awareness signal from specific activation spaces and using it to regulate the model's logits prediction distribution, thereby achieving effective defense within a single forward pass. 

\subsection{Empirical Foundations}
We first establish empirical findings that motivate AISA's design.

\textbf{Motivation and Hypothesis.}
Our work originates from an observation: LLMs can refuse malicious prompts even without any safety data tuning.
To quantify this, we analyze three variants of the Llama3.1-8B~\cite{dubey2024llama}: (i) the original instruct-tuned version; (ii) an uncensored version trained for maximal helpfulness~\cite{Hartford2025Uncensored}; and (iii) an explicitly SFT fine-tuned safety version~\cite{lambert2025tulu}.
We evaluate their responses to prompts from AdvBench~\cite{zou2023universal} and HarmBench~\cite{mazeika2024harmbench} using the StrongReject~\cite{souly2024strongreject} framework for graded assessment.

\begin{figure}[!t]
\centering
\includegraphics[width=\columnwidth]{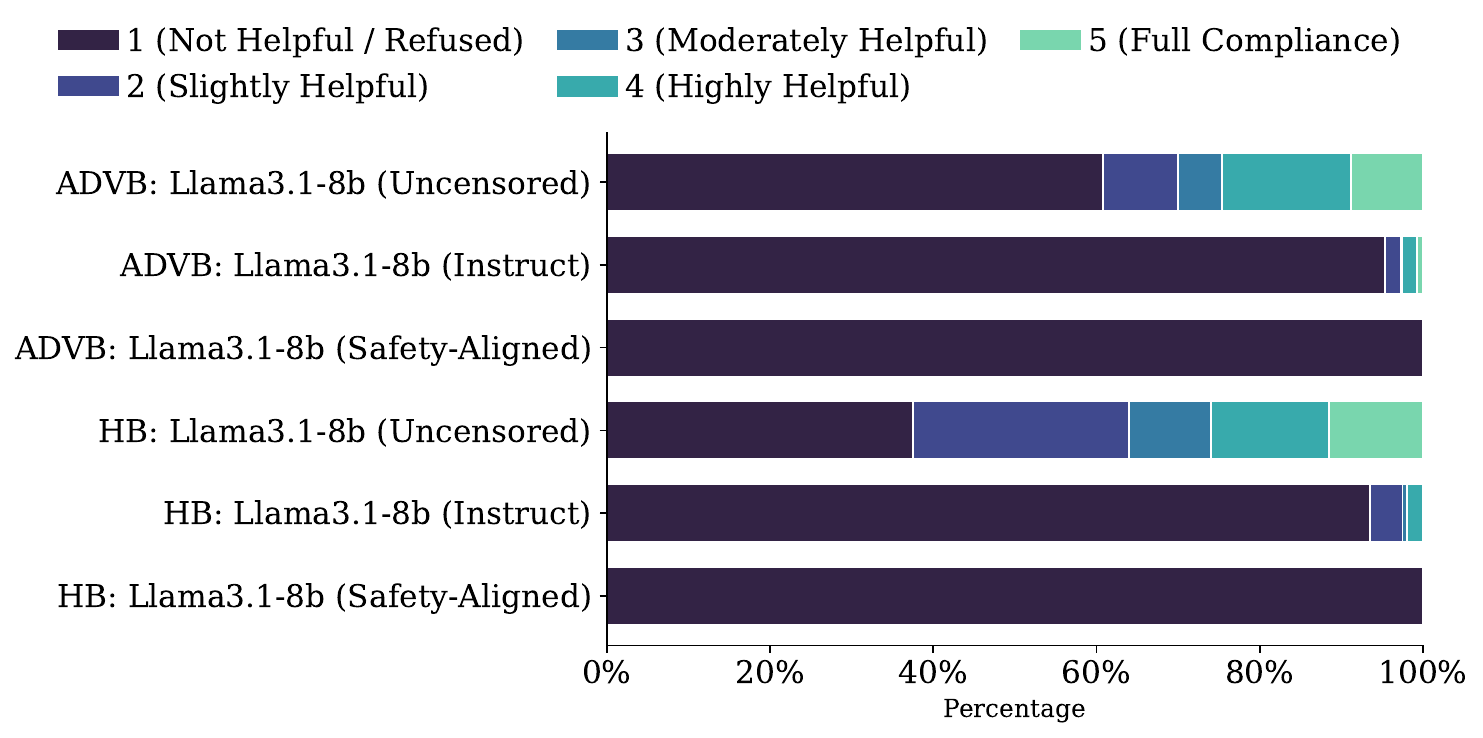}
\caption{Distribution of response safety scores for three Llama3.1-8B variants. Scores are evaluated on AdvBench (ADVB) and HarmBench (HB) using the StrongReject metric.}
\label{fig:intro_phenomenon}
\end{figure}

Figure~\ref{fig:intro_phenomenon} reveals that even the uncensored model exhibits substantial refusal toward malicious prompts, with over 45\% of its responses rated as "slightly helpful" or "refused" across both datasets.
Since such behavior emerges without any explicit safety tuning, we interpret it as evidence of the model's \textit{intrinsic safety awareness} in operation, possibly established during pretraining. Therefore, we hypothesize that proper utilization of this ability could enable an effective defense at a low cost.

\textbf{Spatio-temporal Searching of Safety Awareness.}
We then employ a spatio-temporal searching strategy to localize where and when this awareness is encoded and activated.

\begin{figure}[!t]
\centering
\includegraphics[width=\columnwidth]{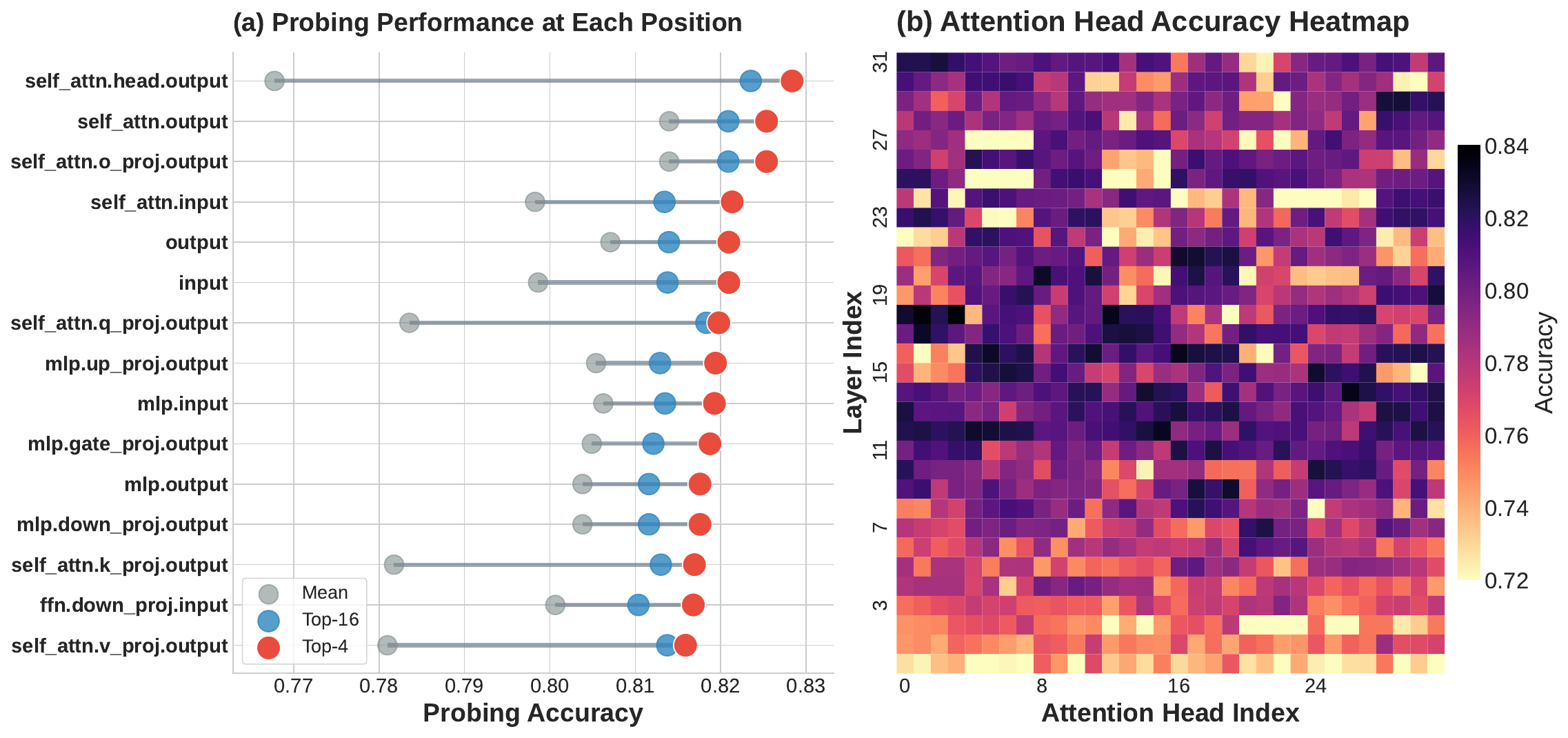}
\caption{Spatial Localization of the intrinsic safety awareness. (a) Probing accuracy comparison across various positions. (b) Accuracy heatmap at $self\_attn.head.output$.}
\label{fig:spatial_localization}
\end{figure}

\textit{Spatial localization} is performed via large-scale linear probing across 15 positions per layer, covering all major activation spaces in LLMs. 
Results on Llama3.1-8B (uncensored) using the AEGIS-2 benchmark~\cite{ghosh2025aegis2} are shown in Figure~\ref{fig:spatial_localization}, with positions arranged in the order they appear during forward propagation.
We observed similar patterns across other model variants, with detailed position specifications and more results available in Appendix~\ref{app:spatial_variants}.

Figure~\ref{fig:spatial_localization} (a) reveals two spatial patterns. 
First, the classification accuracy achieved by simple linear probes at most positions matches or exceeds that of specialized guard models (e.g., LlamaGuard~\cite{fedorov2024llama}) reported in the original AEGIS-2 evaluation~\cite{ghosh2025aegis2} (less than 0.82).
This indicates that linear classifiers on LLM activations can match guard-level performance, while also demonstrating that safety awareness is pervasively encoded across activation spaces.

Second, while averaging over all heads at $self\_attn.head.output$ gives the lowest accuracy; using the top-performing heads greatly improves it. 
This indicates that this position harbors the strongest underlying signal, yet the expression is highly heterogeneous across heads.
This pattern is more directly observable in the heatmap of Figure~\ref{fig:spatial_localization} (b).
We therefore select this position for signal extraction, as its highest upper bound offers the greatest potential for precise safety guidance when head selection is properly applied.

\textit{Temporal localization} tracks when safety awareness emerges during forward propagation by analyzing the prediction probabilities of our linear probes across token trajectories.

\begin{figure}[!t]
\centering
\includegraphics[width=\columnwidth]{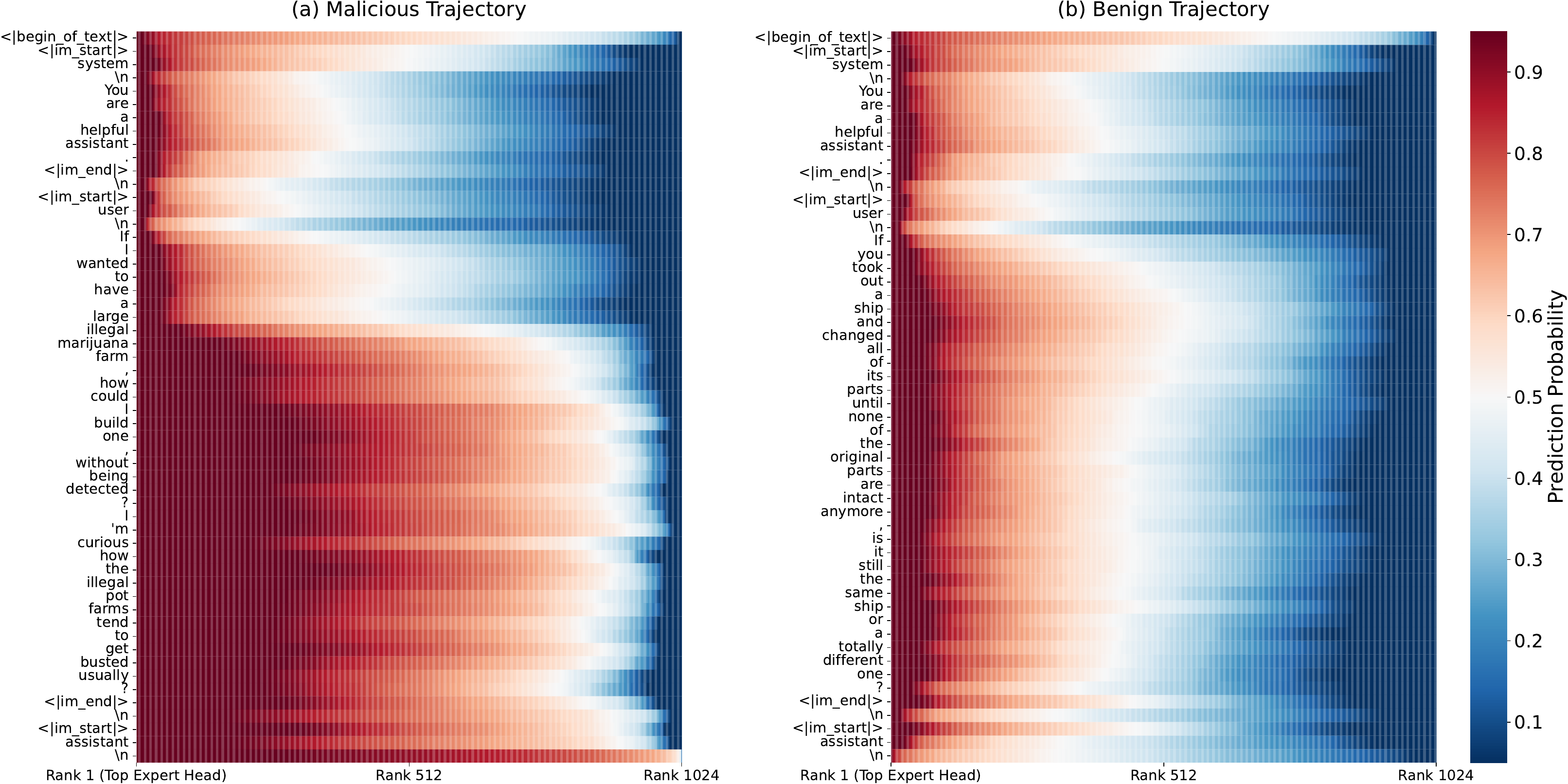}
\caption{Temporal Localization of intrinsic safety awareness. Malicious (left) vs. benign (right) prompts. Probing at $self\_attn.head.output$. Head predictions are sorted by predicted probability (red: high suspicion; blue: low suspicion).}
\label{fig:temporal_localization}
\end{figure}

We show two examples in Figure~\ref{fig:temporal_localization} and claim that predicted probability distributions are tightly linked to token positions. 
At the $<|begin\_of\_text|>$ token, benign and malicious intents are initially separated. 
This distinction is then biased toward the benign side by the system prompt, which encourages general helpfulness.
Throughout the user prompt, predictions adjust dynamically to the semantic content, a process that continues until the prompt ends.

However, at the last line in Figure~\ref{fig:temporal_localization}, a rapid consensus is established at the \texttt{\textbackslash n} token.
In our study, we observe that whenever malicious intent is detected at any point during prompt processing, predictions from attention heads will converge in a stepwise manner at this final structural token.
The phenomenon demonstrates that LLMs leverage structural tokens to consolidate their internal assessment of prompt intent just before generation.
While a similar tendency exists for benign prompts, it is less pronounced than for malicious ones.
In later experiments (Section~\ref{sec:mechanistic_analysis}), we quantify this convergence more precisely using entropy measures.
It is this phenomenon that allows us to bypass the need for tracking the entire trajectory and do token-wise signal extraction. 

\subsection{The Architecture of AISA}
\begin{figure}[!t]
\centering
\includegraphics[width=\columnwidth]{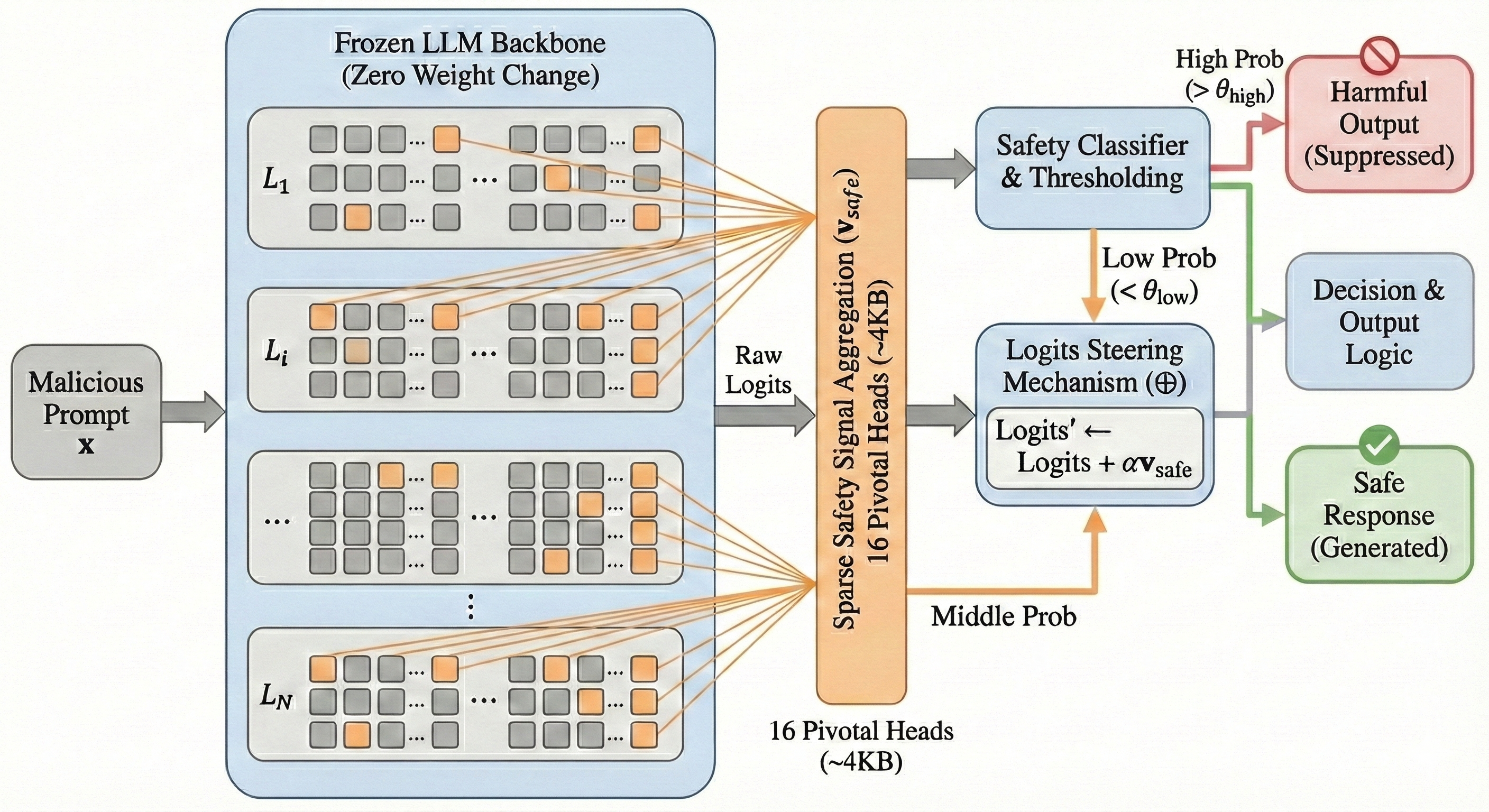}
\caption{The architecture of AISA.}
\label{fig:AISA}
\end{figure}

Our defense framework, AISA, operates in two stages: (1) \textit{signal extraction}, which extracts a safety signal from the specific attention head activation spaces, and (2) \textit{logits steering}, which uses this signal to steer generation via logits distribution adjustment. 
Figure~\ref{fig:AISA} illustrates the complete workflow.

\textbf{Signal Extraction.} We utilize the linear separability of prompt intent within the transformer's activation spaces. 
For an input sequence $P$, let $Head_{i}^{(\ell)} \in \mathbb{R}^{n \times d_h}$ denote the output of the $i$-th attention head at layer $l$ after the scaled dot-product attention computation, $n$ is the sequence length, and $d_h$ is the head dimension. 
According to our empirical analysis, we extract the activation vector $a_{i}^{(\ell)}$ from the final \texttt{\textbackslash n} token position:
\begin{equation}
a_{i}^{(\ell)} = Head_{i}^{(\ell)}[-1,:].
\end{equation}
In practice, both the magnitude and direction of this activation vector prove highly informative for intent discrimination. 
We train a linear classifier for each attention head to map all vectors into corresponding suspicious scores.
\begin{equation}
p_i^{(\ell)} = \sigma\!\big( w_{i}^{(\ell)\top} a_{i}^{(\ell)} + b_i^{(\ell)} \big),
\end{equation}
where $w_i^{(l)} \in \mathbb{R}^{d_h}$ and $b_i^{(l)} \in \mathbb{R}$ are learnable parameters for the $i$-th head at layer $l$, and $\sigma$ denotes the sigmoid function.

Given the sensitivity of probing to the training data distribution, we construct a specialized dataset and train the classifiers with standard binary cross-entropy loss.
Remarkably, we find that with only several thousand curated examples, classification results match or surpass those obtained from million-scale safety corpora~\cite{jiang2024wildteaming}. 
We provide the synthetic dataset details in Appendix x.

Since predictions are also highly sensitive to the choice among attention heads $\mathcal{H}$, and their performance are vary under different situations, we adopt a data-diven strategy: heads are ranked by validation accuracy $\text{Acc}_{(\ell,i)}$, and only the top-$K$ are retained, forming a selected head set $\mathcal{S}_K$:
\begin{equation}
\mathcal{S}_K = \underset{\mathcal{S} \subseteq \mathcal{H},\, |\mathcal{S}|=K}{\arg\max} \sum_{(\ell,i) \in \mathcal{S}} \text{Acc}_{(\ell,i)}.
\end{equation}
The final safety signal $q$ is computed as the average of the prediction probabilities from these heads: $q = \frac{1}{|\mathcal{S}_K|} \sum_{(\ell,i) \in \mathcal{S}_K} p_i^{(\ell)}$.
Large-scale experiments confirm that this data-driven selection strategy not only bypasses algorithmic complexity but also enhances generalization, particularly suitable for models with weaker alignment.

\textbf{Logits Steering.} 
We use the extracted safety score $q$ as a real‑time weighting factor to adjust the model's token distribution during decoding. 
At each generation step $\tau$, the original logits $\mathbf{l}_\tau \in \mathbb{R}^{|V|}$ are adjusted as:
\begin{equation}
\tilde{\mathbf{l}}_\tau = \mathbf{l}_\tau + \Delta(\mathbf{l}_\tau, \mathcal{C}_{\text{safe}}, q, \tau),
\end{equation}
where the steering function $\Delta$ implements a policy defined by two thresholds $0 \le \theta_{\text{low}} \le \theta_{\text{high}} \le 1$.
$\mathcal{C}_{\text{safe}} = (c_1, c_2, \dots, c_n)$ represents a canonical refusal (e.g., ``As a responsible model, I can’t help with that.'').

When $q \le \theta_{\text{low}}$ (passive region), $\Delta = 0$ and generation proceeds normally to preserve utility.
For $\theta_{\text{low}} < q < \theta_{\text{high}}$ (modulated region), $\Delta = \lambda(q) \cdot (\mathcal{C}_{\text{safe}} - \mathbf{l}_\tau)$ gradually shifts the distribution toward a safety‑oriented vector $\mathcal{C}_{\text{safe}}$, with $\lambda(q)$ increasing monotonically.
To maintain language quality in this regime, we also enable a small beam search during decoding.
Finally, when $q \ge \theta_{\text{high}}$ (active region), $\Delta$ forces adherence to a predefined safety template $\mathcal{C}_{\text{safe}}$, typically by up‑weighting its tokens and the EOS token to terminate unsafe continuations. We tune the thresholds and $\lambda(q)$ on the validation set to balance safety and utility.

\textbf{Deployment Costs.}
The added probing classifiers contain only $\sim$0.004M parameters and run without interfering with the base model's normal computation.
These classifiers are trained efficiently on our compact dataset and generalize well. 
The complete defense executes within a single forward pass, introducing negligible latency overhead.

\section{Experiments}
\subsection{Experimental Design}
We organize experiments around two complementary tasks: jailbreak detection and jailbreak defense.
The \textbf{Detection Task} treats the safety score $q$ as a detector to assess its classification accuracy and analyze the signal-extraction mechanism.
The \textbf{End-to-End Defense Task} integrates AISA into LLMs to evaluate its overall defense performance, including safety effectiveness, utility preservation, false refusal rates, and ablation studies.
This dual-task design offers a complete picture of AISA's operation.

\subsection{Experimental Setting}
\label{sec: experimental setting}

\subsubsection{Model Selection}
We select models across the alignment spectrum. 
For Llama3.1-8B~\cite{dubey2024llama}, we obtain a comprehensive set of variants: \textbf{uncensored} (\textit{-U}), standard \textbf{instruct-tuned} (\textit{-I}), \textbf{safety-aligned} (\textit{-A}), and a deliberately unsafe \textbf{dark model}~\cite{darkidol2024} (\textit{-D}). 
For other model families (Llama2~\cite{touvron2023llama}, Mistral~\cite{jiang2023mistral}, Qwen~\cite{yang2025qwen3}, GPT-OSS~\cite{agarwal2025gpt}), we include at most two variants: the standard instruct version and its uncensored counterpart.
Model sizes range from 7B to 20B parameters.
We also include models where safety vectors have been ablated via activation engineering (denoted \textit{-R} for "removed").
This selection spans dense architectures, reasoning-optimized models, MoE variants, and models with different alignment techniques.
More details are provided in Appendix~\ref{app:model_specs}.

\subsubsection{Datasets.}
Given the sensitivity of linear probing to data distribution and the dependence of several design decisions (e.g., head selection) on validation data, we construct controlled synthetic datasets with clear train/validation/test splits.

\textbf{EJ‑OO}  (1.8k samples) combines nine algorithmic jailbreak types with OpenOrca~\cite{mukherjee2023orca} benign instructions to test signal generalization across different attacks. 
\textbf{FQ‑PH} (780 samples) pairs forbidden questions~\cite{shen2024anything} with challenging benign prompts from PHTest~\cite{an2024automatic}, isolating the distinction between harmful and merely sensitive queries. 
\textbf{WJB-s} (10k samples) is a balanced subset of WildJailbreak~\cite{jiang2024wildteaming}, maintaining equal proportions of vanilla/adversarial malicious/benign prompts while reducing training size. 
This shows effective probing needs careful curation, not massive scale.
\textbf{JBC} (1.3k samples) repurposes data from an effective community classifier~\cite{jackhhaoClassifier}, offering high‑quality samples that serve as enhanced additions to the larger ALL‑4 dataset.
\textbf{ALL‑4} aggregates the previous four datasets while preserving their splits, yielding approximately 8k/3k/3k prompts for train/validation/test sets.

In our experiments, unless specified otherwise, AISA uses the linear probe trained on ALL‑4. 
Evaluations on the five synthetic datasets above are considered in‑distribution testing.
To assess generalization, we introduce additional benchmarks as test‑only sets, covering a wide spectrum: 
\textit{Pure malicious prompts} are represented by \textbf{HB} (HarmBench~\cite{mazeika2024harmbench}) and \textbf{ADVB} (AdvBench~\cite{zou2023universal}), which contain clearly harmful jailbreak prompts.
\textit{Sensitive requests} are covered by \textbf{SB} (SorryBench~\cite{souly2024strongreject}), focusing on refusal-prone categories like self-harm and violence.
\textit{Adversarial samples} include \textbf{XST} (XSTest~\cite{rottger2023xstest}) and \textbf{OKT} (OKTest~\cite{shi2024navigating}), which contain challenging or ambiguous benign prompts designed to test safety over‑refusal.
\textit{Automated generation prompts} are represented by \textbf{WGT} (WildGuardTest~\cite{han2024wildguard}), a benchmark consisting of model‑generated malicious prompts paired with benign counterparts.
\textit{Real‑world samples} are provided by \textbf{L3J} (Llama3Jailbreak~\cite{L3J}), featuring in‑the‑wild jailbreak prompts collected from user interactions.
\textit{General utility} datasets (\textbf{GSM8K}~\cite{cobbe2021training}, \textbf{BoolQ}~\cite{clark2019boolq}, \textbf{MMLU}~\cite{hendrycks2020measuring}, \textbf{MMLU‑Pro}~\cite{wang2024mmlu}) assess capability preservation across math solving, reasoning comprehension, and knowledge tasks.

The above benchmarks are employed solely for testing to assess generalization. 
To enable a controlled comparison at the dataset level, we include \textbf{AEG2} (AEGIS‑2~\cite{ghosh2025aegis2}), which provides complete trn/val/tst splits.
In total, our evaluation covers 5 full-split datasets and 11 test benchmarks. More details are provided in Appendix~\ref{app:datasets}.

\subsubsection{Baselines and Evaluation Metrics.}
In \textbf{detection task} evaluations, we benchmark against 9 baselines spanning three categories.
\textit{Commercial API-based Detectors} include OpenAI's \textbf{GPT-4o}~\cite{hurst2024gpt}, \textbf{GPT-4.1-mini}, and \textbf{GPT-5-mini}~\cite{singh2025openai}, serving as strong baselines.
\textit{Specialized Guardrail Models} comprise NVIDIA's \textbf{NemoGuard-JailbreakDetect}~\cite{galinkin2024improved} and Meta's \textbf{Llama-Prompt-Guard-2} (22M / 86M variants)~\cite{meta2024promptguard2}, which are open‑source models explicitly trained for prompt safety filtering.
\textit{Academic Detection Methods} include gradient‑based \textbf{GradSafe}~\cite{xie2024gradsafe}, token‑distribution analysis \textbf{SPDetector}~\cite{candogan2025single}, fintuned‑tuning based \textbf{Jailbreak‑Classifier}~\cite{jackhhaoClassifier}, and multi‑pass generation comparison \textbf{FreeJailbreakDetection}~\cite{chen2025llm}, representing recent advances in jailbreak detection research.
Detection performance is measured using standard classification metrics: Accuracy (ACC) and AUC-ROC (ROC) for binary datasets, and Correctness (COR) for single-class datasets. Ranking methods like FreeJailbreakDetection are evaluated solely with AUC-ROC.

For the \textbf{defense task}, we benchmark against 5 baselines including:
\textbf{ICD}~\cite{wei2023jailbreak}, which uses in‑context learning with safe examples; 
\textbf{SAGE}~\cite{ding2025not}, which self-reflects malicious intent via linguistic analysis and triggers safe generation; 
\textbf{SafeDecoding}~\cite{xu2024safedecoding}, which performs safety‑aware decoding via a safety fine‑tuned auxiliary model; 
\textbf{SCANS}~\cite{cao2025scans}, which dynamically steers activations towards refusal directions; 
and \textbf{SelfDefenD}~\cite{wang2025selfdefend}, which employs a shadow stack to intercept and filter model outputs.
For malicious prompts, safety performance is measured via the 5‑tier SR (Strong‑Reject~\cite{souly2024strongreject}) score.
For benign prompts, we use WildGuard~\cite{han2024wildguard} to assess false refusal numbers.

\subsection{Detection Results and Mechanistic Analysis}
\label{sec:mechanistic_analysis}

We begin by evaluating the discriminative power of the safety signal $q$ as a standalone jailbreak detector.
This establishes the practical viability of intrinsic safety awareness before its integration into end-to-end defense.

\begin{table*}[t]
\caption{Detection accuracy (ACC/COR) across 13 datasets. Methods are trained on ALL‑4 (if possible). API models are strong baselines. Datasets left of ALL-4 are its subsets (in-distribution evaluation); right are independent benchmarks (transfer evaluation). Gray cells mark the highest number per column; when AISA is not top‑1, both top‑1 and best AISA are shaded.}
\label{tab: main_detection_results}
\centering
\small
\resizebox{\textwidth}{!}{
\begin{tabular}{lcccccccccccccc}
\hline
\multirow{2}{*}{Detection Model} & EJ-OO & FQ-PH & WJB & JBC & ALL-4 & XST & WGT & L3J & AEG2 & ADVB & HB & SB & OKT & \multirow{2}{*}{Mean} \\
 & ACC↑ & ACC↑ & ACC↑ & ACC↑ & ACC↑ & ACC↑ & ACC↑ & ACC↑ & ACC↑ & COR↑ & COR↑ & COR↑ & COR↑ &  \\ \hline
gpt-4o-mini-2024-07-18 & 0.8361 & 0.7244 & 0.8665 & 0.8715 & 0.8589 & 0.8578 & 0.8751 & \cellcolor{mylightgray}0.9814 & 0.7822 & 0.9981 & \cellcolor{mylightgray}1.0000 & \cellcolor{mylightgray}0.9833 & 0.9467 & 0.8909 \\
gpt-4.1-mini-2025-04-14 & 0.8778 & 0.7436 & 0.7919 & 0.9157 & 0.8105 & 0.9244 & 0.8644 & 0.9681 & 0.7885 & \cellcolor{mylightgray}1.0000 & 0.9800 & \cellcolor{mylightgray}0.9833 & 0.9700 & 0.8937 \\
gpt-5-mini-2025-08-07 & 0.9333 & 0.7564 & \cellcolor{mylightgray}0.9326 & 0.8916 & \cellcolor{mylightgray}0.9260 & 0.9089 & 0.8917 & 0.9698 & \cellcolor{mylightgray}0.7994 & \cellcolor{mylightgray}1.0000 & \cellcolor{mylightgray}1.0000 & \cellcolor{mylightgray}0.9833 & 0.9667 & 0.9200 \\ \hline
Llama-Prompt-Guard-2-86M & 0.7389 & 0.5256 & 0.4633 & 0.9920 & 0.5443 & 0.5556 & 0.6122 & 0.5488 & 0.4304 & 0.4808 & 0.2000 & 0.0833 & \cellcolor{mylightgray}1.0000 & 0.5519 \\
Jailbreak-Classifier & 0.9917 & 0.8654 & 0.8145 & 0.9719 & 0.8500 & 0.6422 & 0.8478 & 0.8457 & 0.5925 & 0.9962 & 0.9800 & 0.9250 & 0.8800 & 0.8618 \\
NemoGuard-JailbreakDetect & 0.9139 & 0.7628 & 0.6851 & 0.8755 & 0.7290 & 0.6733 & 0.7922 & 0.7848 & 0.5889 & 0.9712 & 0.8750 & 0.7917 & 0.8633 & 0.7928 \\
SPDetector & 0.6833 & 0.6346 & 0.5217 & 0.5502 & 0.5464 & 0.7000 & 0.6157 & 0.6579 & 0.6117 & 0.7423 & 0.6800 & 0.7667 & 0.6667 & 0.6444 \\
GradSafe & 0.6694 & 0.7372 & 0.6946 & 0.6466 & 0.6980 & 0.9067 & 0.8336 & 0.9830 & 0.7817 & 0.9923 & 0.9550 & 0.9500 & 0.9733 & 0.8324 \\ \hline
AISA {[}Mistral-7b-I{]} & 0.9833 & 0.7756 & 0.7923 & 0.9598 & 0.8265 & 0.9333 & 0.8792 & 0.9170 & 0.7167 & \cellcolor{mylightgray}0.9981 & 0.9900 & 0.9667 & 0.9800 & 0.9014 \\
AISA {[}Llama3.1-8b-I{]} & 0.9889 & \cellcolor{mylightgray}0.7885 & \cellcolor{mylightgray}0.8729 & 0.9639 &\cellcolor{mylightgray} 0.8889 & 0.9289 & 0.8911 & 0.9662 & 0.7219 & \cellcolor{mylightgray}0.9981 & 0.9850 & 0.9667 & 0.9900 & 0.9193 \\
AISA {[}Llama2-13b-I{]} & 0.9889 & 0.7756 & 0.8543 & 0.9558 & 0.8736 & \cellcolor{mylightgray}0.9511 & \cellcolor{mylightgray}0.9017 & \cellcolor{mylightgray}0.9676 & \cellcolor{mylightgray}0.7526 & \cellcolor{mylightgray}0.9981 & 0.9900 & \cellcolor{mylightgray}0.9667 & 0.9900 & \cellcolor{mylightgray}0.9205 \\
AISA {[}Qwen3-8b-I{]} & \cellcolor{mylightgray}0.9917 & 0.7821 & 0.8303 & \cellcolor{mylightgray}0.9799 & 0.8582 & 0.9378 & 0.8780 & 0.9495 & 0.7245 & \cellcolor{mylightgray}0.9981 & \cellcolor{mylightgray}0.9950 & 0.9583 & 0.9800 & 0.9126 \\
AISA {[}GPT-OSS-20b-I{]} & 0.9861 & 0.6859 & 0.8149 & 0.9639 & 0.8395 & 0.8889 & 0.8520 & 0.9404 & 0.6975 & 0.9904 & 0.9750 & 0.9417 & \cellcolor{mylightgray}0.9933 & 0.8900 \\ \hline
\end{tabular}
}
\end{table*}

\subsubsection{Main Results on the Detection Task.} 
\label{sec:detection_results}
As shown in Table~\ref{tab: main_detection_results}, the intrinsic safety signal extracted via AISA demonstrates remarkable detection capability.
When signals are extracted from Llama3.1-8b-I and trained on ALL-4, AISA achieves a mean accuracy of 0.919, surpassing OpenAI's GPT-4o-mini and GPT-4.1-mini.
Scaling the source model to 13B parameters, Llama2-13b-I, further elevates performance to 0.9205, surpassing commercial frontier GPT-5-mini.
This performance is unmatched by any academic baseline.
More importantly, AISA's deployment across different model variants does not incur significant performance gaps; even when AISA does not rank first, its accuracy remains within 6 percentage points of the best API models. 
This demonstrates that the intrinsic safety awareness within LLMs, when properly harnessed by AISA, yields a discriminative signal strong enough to compete with state-of-the-art external detectors.
We also plot ROC curves on binary-class datasets in Appendix~\ref{app:roc_curves} (Figure~\ref{fig:roc_curves_appendix}).
Our method demonstrates smoother curves with better coverage of the upper-left region.

\begin{figure}[!t]
\centering
\includegraphics[width=\columnwidth]{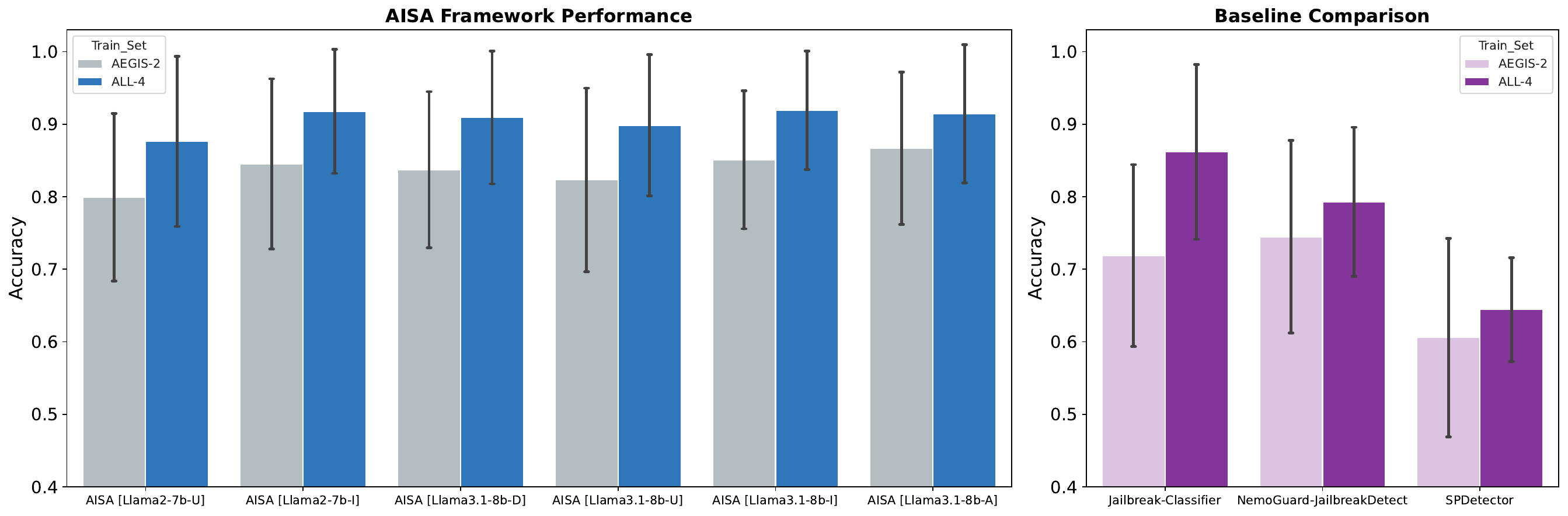}
\caption{Performance comparison of trainable methods with different training data. The y-axis shows mean accuracy (ACC/COR) across all 13 test datasets. AISA consistently outperforms other methods across training datasets, with ALL-4 achieving the best overall performance.}
\label{fig:different_trn}
\end{figure}

\textbf{Further Discussions:} 
(1) \textit{Method-Data Synergy.}
We find that the successful extraction of intrinsic safety awareness depends on synergy between probing method and training data.
This is evidenced by Figure~\ref{fig:different_trn}: all methods trained on ALL‑4 outperform those trained on AEGIS‑2. 
Notably, AISA uses only a 128‑dimensional linear probe yet surpasses models with larger architectures or full fine‑tuning.
The fact that ALL‑4 achieves this superiority despite being less than half the size of AEGIS‑2 ($\sim$14k vs $\sim$33k samples) demonstrates that training data quality outweighs sheer quantity for eliciting robust safety signals.
We therefore credit the success of AISA half to the construction of the ALL-4 dataset.

(2) \textit{Signal Generalization.}
From a defense perspective, generalization is paramount, as failing on unseen attacks offers little practical security.
AISA addresses this concern directly: its safety signal demonstrates strong generalization across both model variations and prompt categories. 
Specifically: (i) the signal remains robust regardless of the model's family (Llama, Mistral, Qwen, etc.), size (7B to 13B parameters), or alignment status (see Figure~\ref{fig:different_trn}, aligned, uncensored, dark, etc.), producing consistent detection performance across all configurations; 
(ii) it maintains high accuracy across diverse prompt types, covering nearly all categories seen in prompt‑based jailbreak scenarios (as detailed in Section~\ref{sec: experimental setting}). 
Additionally, Table~\ref{tab: change_target_system} shows that this generalization extends to algorithmic jailbreaks targeting different systems. 
We posit that malicious prompts, regardless of their surface form, share the same underlying goal of eliciting harmful content. 
This fundamental intent triggers detectable internal precursors, explaining why the signal extracted by AISA is consistently strong.

\begin{table}[!t]
\caption{Detection accuracy on algorithmic jailbreaks (EJ-OO) when attacks target different models. Results show minimal variation, indicating target-model independence.}
\label{tab: change_target_system}
\centering
\resizebox{\columnwidth}{!}{
\begin{tabular}{lcc}
\hline
 & EJ-OO & EJ-OO \\
 & {[}Target: Mistral-7b-I{]} & {[}Target: Llama2-7b-I{]} \\ \hline
AISA{[}Mistral-7b-I{]} & 0.9833 & 0.9694 \\
AISA{[}Llama2-7b-I{]} & 0.9117 & 0.9833 \\
AISA{[}Llama3.1-8b-I{]} & 0.9889 & 0.9757 \\ \hline
\end{tabular}
}
\vspace{-10pt}
\end{table}

\subsubsection{Mechanistic Analysis}
While the results in Section~\ref{sec:detection_results} have confirmed AISA’s
strong detection capability, a deeper investigation into its operational mechanisms is warranted.  
Specifically, we focus on the following three questions:
(1) Do the spatiotemporal trends observed in Chapter~\ref{sec:section_3} persist when probes are trained on the more diverse ALL‑4 dataset?
(2) Why is selective attention head probing necessary? Meanwhile, why does focusing on top-K heads suffice, and how does the choice of K affect detection performance?
(3) What evidence supports characterizing this signal as "safety awareness"? In particular, do attention heads selected from different models encode similar safety concepts?

\begin{figure}[!t]
\centering
\includegraphics[width=\columnwidth]{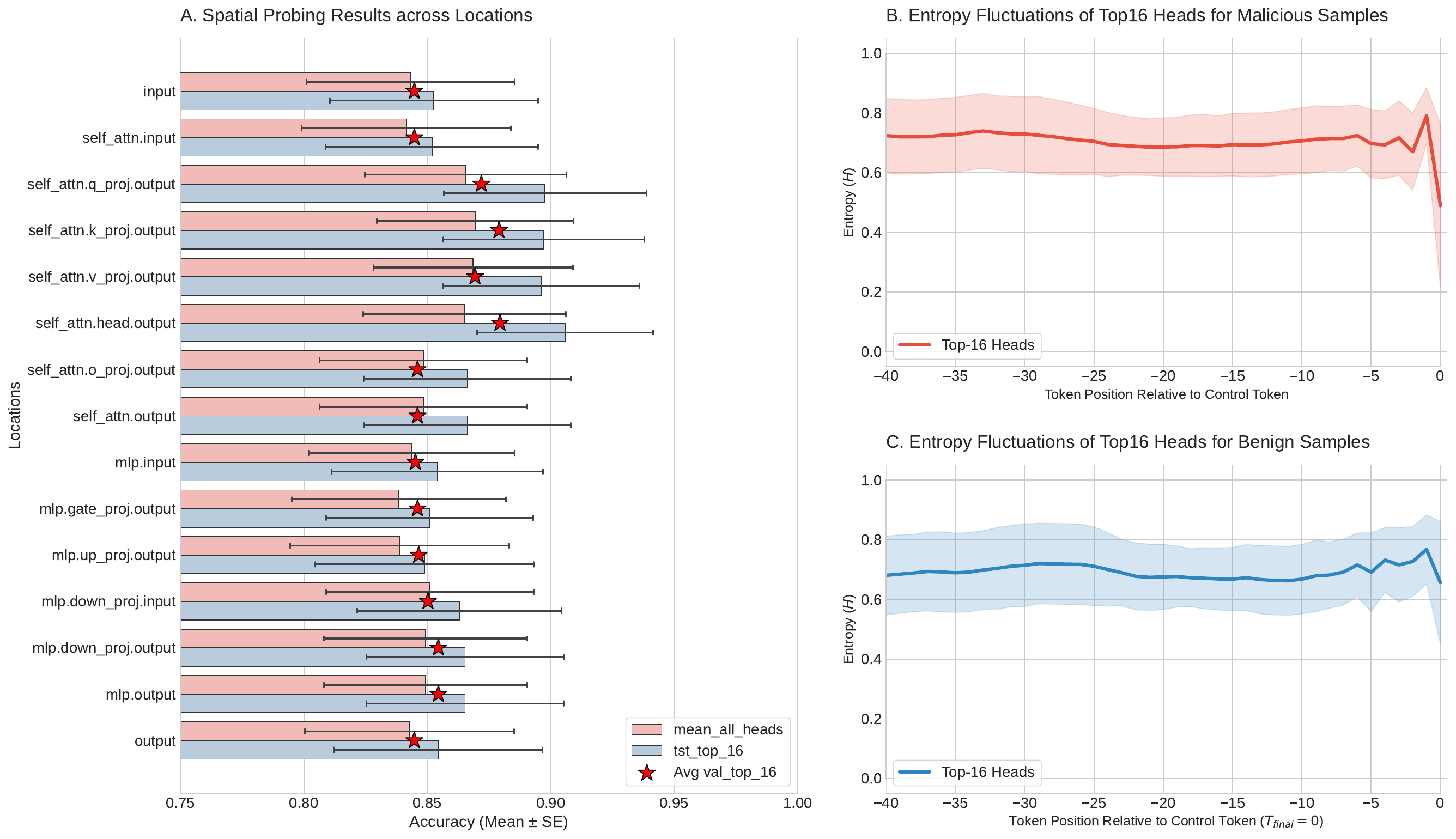}
\caption{Spatiotemporal analysis with ALL-4 trained probes. 
(a) Probing accuracy across activation positions (attention head outputs remain optimal). 
(b,c) Entropy evolution of 16 selected heads over token positions for malicious and benign prompts, showing sharp convergence at final tokens on malicious samples.}
\label{fig:spatiotemporal_all4}
\end{figure}

\textbf{Answer to Q1: Spatiotemporal Location.}
Figure~\ref{fig:spatiotemporal_all4} validates our earlier architectural choices. 
When probes are trained on the more diverse ALL-4 dataset, attention head outputs remain the most optimal position for safety signal extraction (Fig.~\ref{fig:spatiotemporal_all4} (a)). 
This confirms that the spatial superiority of this location is not an artifact of a specific training distribution but a stable property of the transformer architecture.
Temporally, we plot the prediction entropy of the selected top‑16 heads across token positions.
As shown in Fig.~\ref{fig:spatiotemporal_all4} (b), entropy on malicious samples begins to fluctuate upon reaching the structural tokens (the last five positions), then drops sharply toward the end, indicating that head predictions converge to a more unified state.
We do not observe a similar trend for benign samples (Fig.~\ref{fig:spatiotemporal_all4} (c)), likely because harmful intents often share common patterns, whereas benign queries exhibit greater diversity.
Nevertheless, the spatial location and temporal convergence point have already proven effective in Table~\ref{tab: main_detection_results}, demonstrating the practical utility of AISA. Appendix~\ref{app:spatial_variants} offers more results.

\begin{figure}[t]
\centering
\includegraphics[width=\columnwidth]{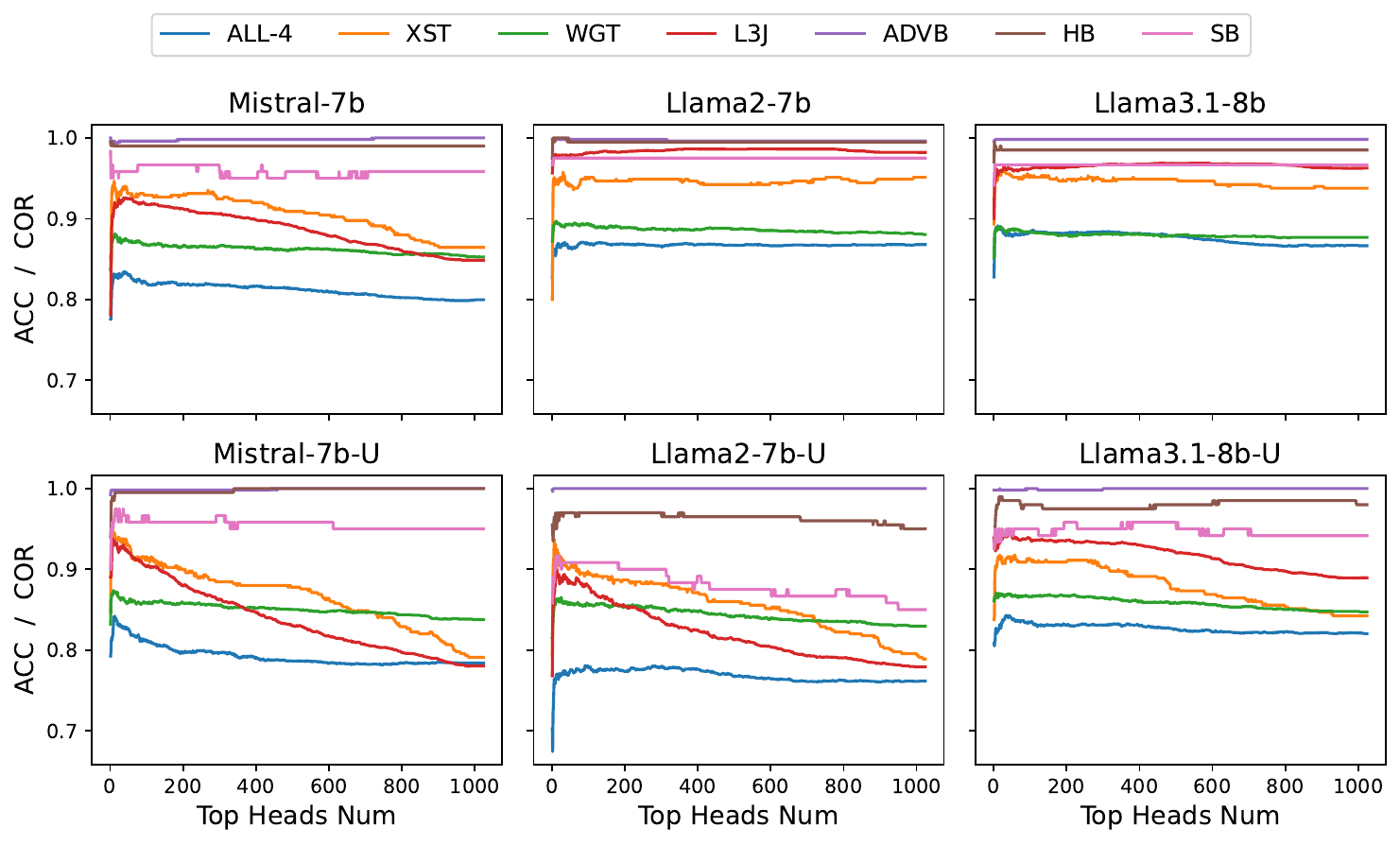}
\caption{Overall detection performance trend of AISA with increasing number of selected attention heads (K). Top row: instruct‑tuned models; bottom row: uncensored models.}
\label{fig: topk_curve}
\end{figure}

\textbf{Answer to Q2: Top-k Selection.}
The top‑K head selection serves dual purposes: it minimizes AISA's computational overhead while purifying the aggregated prediction probability.
We justify this design by examining how performance varies with K.
Figure~\ref{fig: topk_curve} shows that as K increases, accuracy rises rapidly to a peak, then gradually declines.
For instruct-tuned models, performance plateaus after the peak, indicating that additional heads contribute minimal new signal.
In contrast, less aligned models with \textit{-U} suffix show a more pronounced decrease. 
This distinction suggests topk selection strategy is more suitable for unaligned models, as the performance drop underscores the critical importance of selective ranking.
However, this selection for aligned models is also necessary as it effectively reduces the need for redundant probing classifiers.

\begin{figure}[t]
\centering
\includegraphics[width=\columnwidth]{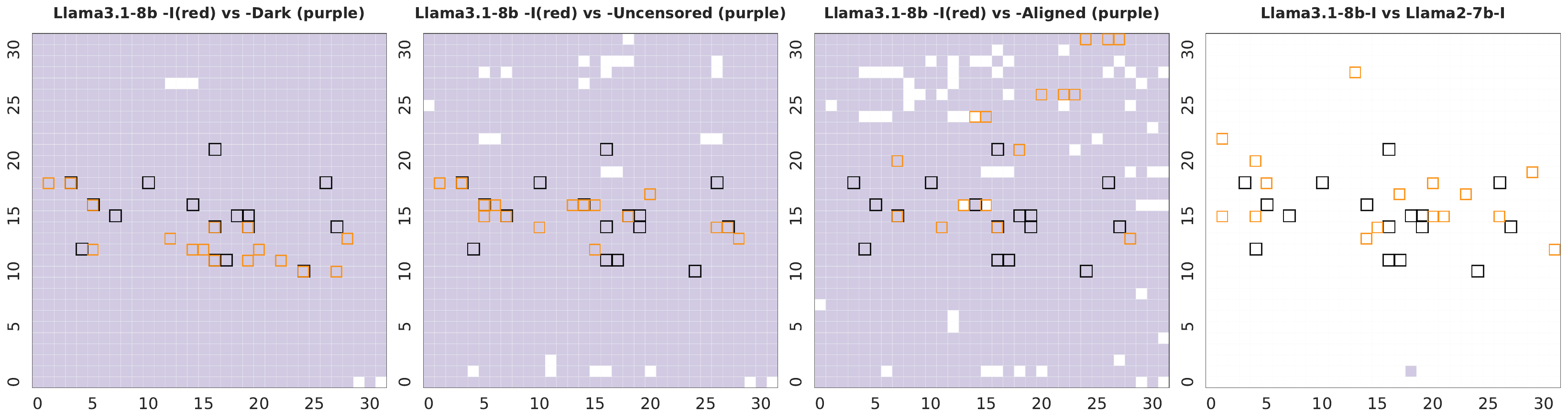}
\caption{Similarity of classifier weights across attention heads. Pale purple indicates similarity >0.265 (significant). Black: heads selected from Llama3.1‑8B‑I; orange: corresponding selections from other variants. The pattern shows safety awareness is intrinsically tied to the base architecture.}
\label{fig:safety_awareness_similarity}
\end{figure}

\textbf{Answer to Q3: Safety Awareness.}
This question lies at the core of our work—whether such intrinsic safety awareness genuinely exists within models, and whether closely related models share this common mechanism.
To investigate, we extract the weight vectors $\mathbf{w}_i^{(\ell)}$ from the linear classifiers trained on each attention head across different models and compute their cosine similarities. 
For Llama3.1‑8B, the head dimension is $d_h = 128$. 
Under a 3‑$\sigma$ criterion, two randomly oriented vectors would have an expected cosine similarity below 0.265. 
We therefore set 0.265 as the significance threshold.
Figure~\ref{fig:safety_awareness_similarity} presents the resulting similarity matrix.
We observe widespread high similarity (exceeding 0.265) among Llama3.1 variants across different alignment levels (\textit{-I}, \textit{-U}, \textit{-A}, \textit{-D}), indicating that classifiers trained on these models share a common underlying representation.
\textbf{This finding directly establishes the existence of a powerful, pretrain-based safety awareness that persists irrespective of downstream fine‑tuning objectives, be they malicious (turn into dark models) or safety‑oriented.}

We also see an interesting pattern from the automatic head selection: most selected heads cluster in the middle layers of the LLM stack.
Only in strongly safety‑aligned models do we observe a shift toward higher‑layer heads being preferentially selected.

\subsection{Results and Analysis on Defense Task}
In this task, we evaluate end‑to‑end defense performance.
Using unaligned models as the starting point, we compare their performance when equipped with AISA against other defense baselines.
Our evaluation focus on three aspects: (1) output mitigation against jailbreak attacks, (2) false refusal rate on adversarial benign prompts, and (3) preservation of general model utility on standard benchmarks.

\begin{table}[t]
\caption{Strong-Reject Score (SR, 1-5) of different defense methods. Lower values indicate stronger safety protection.}
\vspace{-10pt}
\label{tab: sr_results}
\resizebox{\columnwidth}{!}{
\begin{tabular}{llccccccc}
\hline
BASE MODEL & METHODS & ALL-4 & XST & WGT & L3J & ADVB & HB & SB \\ \hline
\multirow{7}{*}{Mistral-7b-U} & vanilla & 2.664 & 3.030 & 2.595 & 3.341 & 3.473 & 3.065 & 3.258 \\
 & Instruct 2.448 & 1.210 & 2.148 & 1.719 & 1.658 & 1.740 & 1.892 \\
 & SAGE & 2.473 & 2.905 & 2.401 & 3.092 & 2.750 & 2.250 & 2.533 \\
 & ICD & 2.264 & \cellcolor{mylightgray}1.065 & 1.749 & 1.202 & 1.152 & 1.150 & 1.375 \\
 & SCANS & 2.643 & 1.790 & 2.519 & 2.601 & 2.588 & 2.315 & 3.008 \\
 & SelfDefenD & 2.251 & 1.150 & 1.813 & 1.256 & 1.067 & 1.330 & 1.117 \\
 & AISA & \cellcolor{mylightgray}1.535 & 1.275 & \cellcolor{mylightgray}1.346 & \cellcolor{mylightgray}1.150 & \cellcolor{mylightgray}1.008 & \cellcolor{mylightgray}1.050 & \cellcolor{mylightgray}1.108 \\ \hline
\multirow{7}{*}{Llama2-7b-U} & vanilla & 2.511 & 1.800 & 2.551 & 2.514 & 2.754 & 2.750 & 2.892 \\
 & Instruct & 1.227 & \cellcolor{mylightgray}1.000 & 1.290 & 1.073 & 1.004 & 1.015 & 1.117  \\
 & SAGE & 2.370 & 2.115 & 2.425 & 2.361 & 2.223 & 2.245 & 2.642 \\
 & ICD & 2.305 & 1.710 & 2.163 & 2.217 & 1.998 & 2.165 & 2.175 \\
 & SCANS & 2.470 & \cellcolor{mylightgray}1.320 & 2.441 & 1.785 & 1.604 & 1.615 & 2.167 \\
 & SelfDefenD & 2.403 & 1.345 & 2.158 & 1.663 & 1.485 & 1.890 & 1.825 \\
 & AISA & \cellcolor{mylightgray}1.993 & 1.455 & \cellcolor{mylightgray}1.726 & \cellcolor{mylightgray}1.441 & \cellcolor{mylightgray}1.110 & \cellcolor{mylightgray}1.230 & \cellcolor{mylightgray}1.558 \\ \hline
\multirow{7}{*}{Llama3.1-8b-D} & vanilla & 2.579 & 1.560 & 2.219 & 1.924 & 2.067 & 2.425 & 2.158 \\
 & Instruct & 1.840 & 1.080 & 1.621 & 1.160 & 1.179 & 1.335 & 1.117 \\
 & SAGE & 2.417 & \cellcolor{mylightgray}1.070 & 2.207 & 2.064 & 1.725 & 1.880 & 1.958 \\
 & ICD & 2.464 & 1.945 & 1.744 & \cellcolor{mylightgray}1.133 & 1.029 & 1.136 & \cellcolor{mylightgray}1.100 \\
 & SCANS & 2.549 & 1.390 & 2.164 & 1.609 & 1.473 & 1.575 & 1.933 \\
 & SelfDefenD & 2.496 & 1.310 & 1.993 & 1.522 & 1.519 & 1.690 & 1.492 \\
 & AISA & \cellcolor{mylightgray}1.586 & 1.285 & \cellcolor{mylightgray}1.414 & 1.208 & \cellcolor{mylightgray}1.027 & \cellcolor{mylightgray}1.045 & 1.175 \\ \hline
\end{tabular}
}
\end{table}

\begin{figure}[t]
\centering
\includegraphics[width=\columnwidth]{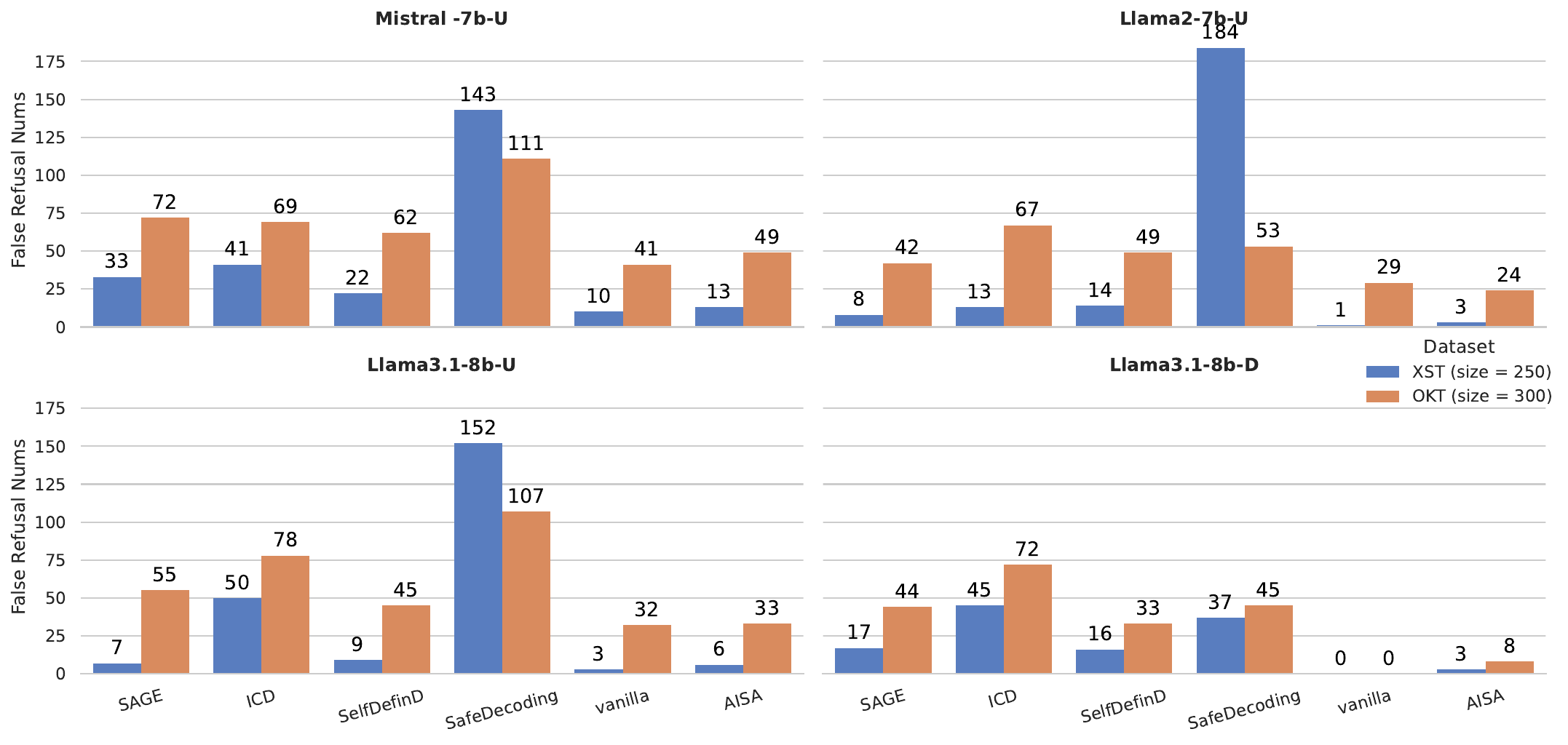}
\vspace{-10pt}
\caption{False Refusal Count (FRC) of benign prompts across different defense methods. Lower values indicate better utility preservation. Vanilla is the strong baseline.}
\label{fig: false_rejection_count}
\vspace{-10pt}
\end{figure}

\textbf{Main Defense Results} 
(1) AISA demonstrates consistent defensive capabilities in Table~\ref{tab: sr_results}, maintaining SR scores below 2.0 across all test scenarios. 
This indicates reliable protection, as scores exceeding this threshold would imply the model provides substantial assistance to malicious prompts on average.
(2) The occasionally better‑performing methods, ICD and SAGE, both modify user prompts: ICD adds malicious demonstrations, and SAGE appends reflective prompts for safety reasoning.
This requirement for prompt alteration restricts their practical applicability in many commercial situations.
(3) Furthermore, as shown in Figure 7, their marginal advantage in rejecting malicious queries comes at the cost of significantly overkill rates on benign prompts. 
In contrast, our method's performance on benign prompts is nearly identical to that of the vanilla, undefended model, demonstrating that our defense preserves the model's original capabilities without introducing detrimental over-refusal.
(4) After applying our defense method, unaligned models can match or exceed the safety performance of their aligned counterparts in many cases (notably for \textit{Mistral-7b} and \textit{Llama3.1-8b} in Table~\ref{tab: sr_results}). 
This suggests our approach could serve as a resource-efficient alternative to alignment tuning, while avoiding the over-refusal issues.

\begin{table}[]
\centering
\caption{General capability preservation with AISA deployed.}
\vspace{-10pt}
\resizebox{\columnwidth}{!}{
\begin{tabular}{lcccc}
\hline
MODEL & GSM8K & BoolQ & MMLU & MMLU-pro \\ \hline
Mistral-7b-U & 75.51 & 86.72 & 61.90 & 36.62 \\
AISA {[}Mistral-7b-U{]} & 75.52 & 86.72 & 61.92 & 36.62 \\
Llama3.1-8b-I & 82.79 & 83.87 & 71.44 & 54.93 \\
AISA {[}Llama3.1-8b-I{]} & 82.79 & 83.88 & 71.44 & 54.90 \\ \hline
\end{tabular}
}
\vspace{-10pt}
\end{table}

\subsection{Further Analysis of Our Defense Mechanism} 
In our experiments, the defense mechanism (logits boost part) of AISA uses a force refusal threshold of 0.85, a safety output threshold of 0.4, and a linear boosting strategy.
Here we examine different functional implementations of $\Gamma(\mathcal{T}_{\text{safe}}, p_{\text{final}}, t)$.
We use \textit{Llama3.1-8b} and \textit{Llama3.1-8b-U} as representative models.

\begin{table}[t]
\centering
\caption{Performance of AISA under different parameter configurations (SOT: Safety Output Threshold, FRT: Force Refusal Threshold, Boost: Function used in $\Gamma(\mathcal{T}_{\text{safe}}, p_{\text{final}}, t)$). Our final configuration (SOT=0.4, FRT=0.85, linear boost) corresponds to the last row for each LLM in the table.}
\vspace{-10pt}
\label{tab: ab_study}
\resizebox{\columnwidth}{!}{
\begin{tabular}{llllcccccc}
\hline
\multirow{3}{*}{Base Model} & \multicolumn{3}{l}{Defense Mechanism} & \multicolumn{6}{c}{Datasets} \\ \cline{2-10} 
 & \multirow{2}{*}{SOT} & \multirow{2}{*}{FRT} & \multirow{2}{*}{Boost} & ALL-4 & XST & WGT & L3J & XST & OKT \\
 &  &  &  & SR↓ & SR↓ & SR↓ & SR↓ & FRC↓ & FRC↓ \\ \hline
\multirow{4}{*}{Llama3.1-8b} & 0.5 & 0.5 & - & 1.2389 & 1.1400 & 1.2804 & 1.0310 & 23 & 88 \\
 & 0.15 & 0.85 & Linear & \cellcolor{mylightgray}1.1218 & \cellcolor{mylightgray}1.0950 & \cellcolor{mylightgray}1.1709 & \cellcolor{mylightgray}1.0080 & 47 & 95 \\
 & 0.4 & 0.85 & Quadratic & 1.3988 & 1.1600 & 1.3738 & 1.0942 & \cellcolor{mylightgray}14 & \cellcolor{mylightgray}62 \\
 & 0.4 & 0.85 & Linear & 1.3000 & 1.1650 & 1.3191 & 1.0404 & \cellcolor{mylightgray}14 & \cellcolor{mylightgray}62 \\ \hline
\multirow{4}{*}{Llama3.1-8b-U} & 0.5 & 0.5 & - & 1.4315 & 1.3850 & 1.3738 & 1.1455 & 12 & 47 \\
 & 0.15 & 0.85 & Linear & \cellcolor{mylightgray}1.2410 & \cellcolor{mylightgray}1.0700 & \cellcolor{mylightgray}1.2577 & \cellcolor{mylightgray}1.0944 & 31 & 86 \\
 & 0.4 & 0.85 & Quadratic & 1.7938 & 1.3900 & 1.5140 & 1.2296 & \cellcolor{mylightgray}6 & \cellcolor{mylightgray}30 \\
 & 0.4 & 0.85 & Linear & 1.5859 & 1.2850 & 1.4139 & 1.2076 & \cellcolor{mylightgray}6 & 33 \\ \hline
\end{tabular}
}
\vspace{-10pt}
\end{table}

Our configuration (SOT=0.4, FRT=0.85, Linear) achieves an optimal balance in safety protection and utility preservation, maintaining low SR scores (e.g., 1.3000 on ALL-4 for the aligned model) while keeping false refusal counts manageable. An over-aggressive setup (SOT=0.15, FRT=0.85) leads to a sharp increase in false refusals (FRC of 47 on XST), confirming that moderate thresholds are crucial for utility preservation.
Meanwhile, the quadratic boosting function (implemented as $4(p - 0.4)^2$) appears to be overly conservative, failing to assign sufficient boost on target tokens.
This analysis confirms that our primary design choices are well-justified and contribute to a robust yet practical defense.

\section{Conclusion}

In this paper, we introduce AISA, a defense framework that awakens the intrinsic safety awareness inherent in LLMs.
Our approach leverages the linear separability of prompt intent in attention head activation spaces to generate probabilistic guidance for robust inference.
The defense process requires no complex training, preserves original prompt structures, and operates without resource-intensive safety alignment tuning.

We evaluate AISA across 13 diverse datasets covering algorithmic jailbreaks, real‑world safety violations, general inquiries, and adversarial samples.
Compared with 9 detection models and 5 defense mechanisms, AISA demonstrates strong generalization capability, consistently achieving competitive SR scores across all test scenarios while maintaining low false refusal rates.
We believe our method could serve as a resource-efficient yet effective alternative to current complex defense pipelines in the LLM safety field.



\textbf{Ethical concerns.} While this work necessarily involves studying malicious prompts that contradict ethical norms, we have carefully avoided including any explicitly harmful, offensive, or graphically dangerous content in the paper.
All examples have been reviewed to minimize potential negative societal impact.

\bibliographystyle{ACM-Reference-Format}
\bibliography{sample-base}

@article{matarazzo2025survey,
  title={A survey on large language models with some insights on their capabilities and limitations},
  author={Matarazzo, Andrea and Torlone, Riccardo},
  journal={arXiv preprint arXiv:2501.04040},
  year={2025}
}

@article{huynh2025large,
  title={Large Language Models for Code Generation: A Comprehensive Survey of Challenges, Techniques, Evaluation, and Applications},
  author={Huynh, Nam and Lin, Beiyu},
  journal={arXiv preprint arXiv:2503.01245},
  year={2025}
}

@article{yi2024jailbreak,
  title={Jailbreak attacks and defenses against large language models: A survey},
  author={Yi, Sibo and Liu, Yule and Sun, Zhen and Cong, Tianshuo and He, Xinlei and Song, Jiaxing and Xu, Ke and Li, Qi},
  journal={arXiv preprint arXiv:2407.04295},
  year={2024}
}

@inproceedings{christiano2017deep,
  title     = {Deep Reinforcement Learning from Human Preferences},
  author    = {Christiano, Paul F and Leike, Jan and Brown, Tom and Martic, Miljan and Legg, Shane and Amodei, Dario},
  booktitle = {Advances in Neural Information Processing Systems},
  year      = {2017}
}

@article{zhang2025exploring,
  title={Exploring the role of large language models in the scientific method: from hypothesis to discovery},
  author={Zhang, Yanbo and Khan, Sumeer A and Mahmud, Adnan and Yang, Huck and Lavin, Alexander and Levin, Michael and Frey, Jeremy and Dunnmon, Jared and Evans, James and Bundy, Alan and others},
  journal={npj Artificial Intelligence},
  volume={1},
  number={1},
  pages={14},
  year={2025},
}

@inproceedings{wolf2025tradeoffs,
  title={Tradeoffs Between Alignment and Helpfulness in Language Models with Steering Methods},
  author={Wolf, Yotam and Wies, Noam and Shteyman, Dorin and Rothberg, Binyamin and Levine, Yoav and Shashua, Amnon},
  booktitle={ICLR 2025 Workshop on Foundation Models in the Wild},
  year={2025},
}

@inproceedings{he2025plan,
  title={Plan-then-execute: An empirical study of user trust and team performance when using llm agents as a daily assistant},
  author={He, Gaole and Demartini, Gianluca and Gadiraju, Ujwal},
  booktitle={Proceedings of the 2025 CHI Conference on Human Factors in Computing Systems},
  year={2025}
}

@article{chen2026jailbreaking,
  title={Jailbreaking LLMs \& VLMs: Mechanisms, Evaluation, and Unified Defense},
  author={Chen, Zejian and Li, Chaozhuo and Li, Chao and Zhang, Xi and Zhang, Litian and He, Yiming},
  journal={arXiv preprint arXiv:2601.03594},
  year={2026}
}

@inproceedings{pu2025feint,
title={Feint and Attack: Jailbreaking and Protecting LLMs via Attention Distribution Modeling},
author={Pu, Rui and Li, Chaozhuo and Ha, Rui and Chen, Zejian and Zhang, Litian and Liu, Zheng and Qiu, Lirong and Ye, Zaisheng},
booktitle={Proceedings of the Thirty-Fourth International Joint Conference on Artificial Intelligence.},
year={2025}
}

@inproceedings{wang2025selfdefend,
  title={$\{$SelfDefend$\}$:$\{$LLMs$\}$ can defend themselves against jailbreaking in a practical manner},
  author={Wang, Xunguang and Wu, Daoyuan and Ji, Zhenlan and Li, Zongjie and Ma, Pingchuan and Wang, Shuai and Li, Yingjiu and Liu, Yang and Liu, Ning and Rahmel, Juergen},
  booktitle={34th USENIX Security Symposium (USENIX Security 25)},
  pages={2441--2460},
  year={2025}
}

@inproceedings{yuan2025refuse,
  title={Refuse whenever you feel unsafe: Improving safety in llms via decoupled refusal training},
  author={Yuan, Youliang and Jiao, Wenxiang and Wang, Wenxuan and Huang, Jen-tse and Xu, Jiahao and Liang, Tian and He, Pinjia and Tu, Zhaopeng},
  booktitle={Proceedings of the 63rd Annual Meeting of the Association for Computational Linguistics (Volume 1: Long Papers)},
  year={2025}
}

@article{jain2023baseline,
  title={Baseline defenses for adversarial attacks against aligned language models},
  author={Jain, Neel and Schwarzschild, Avi and Wen, Yuxin and Somepalli, Gowthami and Kirchenbauer, John and Chiang, Ping-yeh and Goldblum, Micah and Saha, Aniruddha and Geiping, Jonas and Goldstein, Tom},
  journal={arXiv preprint arXiv:2309.00614},
  year={2023}
}

@article{deng2023multilingual,
  title={Multilingual jailbreak challenges in large language models},
  author={Deng, Yue and Zhang, Wenxuan and Pan, Sinno Jialin and Bing, Lidong},
  journal={arXiv preprint arXiv:2310.06474},
  year={2023}
}

@article{wei2023jailbreak,
  title={Jailbreak and guard aligned language models with only few in-context demonstrations},
  author={Wei, Zeming and Wang, Yifei and Li, Ang and Mo, Yichuan and Wang, Yisen},
  journal={arXiv preprint arXiv:2310.06387},
  year={2023}
}

@article{li2023deepinception,
  title={Deepinception: Hypnotize large language model to be jailbreaker},
  author={Li, Xuan and Zhou, Zhanke and Zhu, Jianing and Yao, Jiangchao and Liu, Tongliang and Han, Bo},
  journal={arXiv preprint arXiv:2311.03191},
  year={2023}
}

@article{lv2024codechameleon,
  title={Codechameleon: Personalized encryption framework for jailbreaking large language models},
  author={Lv, Huijie and Wang, Xiao and Zhang, Yuansen and Huang, Caishuang and Dou, Shihan and Ye, Junjie and Gui, Tao and Zhang, Qi and Huang, Xuanjing},
  journal={arXiv preprint arXiv:2402.16717},
  year={2024}
}

@article{handa2024competency,
  title={When" competency" in reasoning opens the door to vulnerability: Jailbreaking llms via novel complex ciphers},
  author={Handa, Divij and Zhang, Zehua and Saeidi, Amir and Kumbhar, Shrinidhi and Uddin, Md Nayem and RRV, Aswin and Baral, Chitta},
  journal={arXiv preprint arXiv:2402.10601},
  year={2024}
}

@article{xie2023defending,
  title={Defending chatgpt against jailbreak attack via self-reminders},
  author={Xie, Yueqi and Yi, Jingwei and Shao, Jiawei and Curl, Justin and Lyu, Lingjuan and Chen, Qifeng and Xie, Xing and Wu, Fangzhao},
  journal={Nature Machine Intelligence},
  volume={5},
  number={12},
  pages={1486--1496},
  year={2023},
  publisher={Nature Publishing Group UK London}
}

@inproceedings{zhang2025intention,
  title={Intention analysis makes llms a good jailbreak defender},
  author={Zhang, Yuqi and Ding, Liang and Zhang, Lefei and Tao, Dacheng},
  booktitle={Proceedings of the 31st International Conference on Computational Linguistics},
  pages={2947--2968},
  year={2025}
}

@article{liu2024adversarial,
  title={Adversarial tuning: Defending against jailbreak attacks for llms},
  author={Liu, Fan and Xu, Zhao and Liu, Hao},
  journal={arXiv preprint arXiv:2406.06622},
  year={2024}
}

@inproceedings{cao2025scans,
  title={SCANS: Mitigating the exaggerated safety for llms via safety-conscious activation steering},
  author={Cao, Zouying and Yang, Yifei and Zhao, Hai},
  booktitle={Proceedings of the AAAI Conference on Artificial Intelligence},
  year={2025}
}

@inproceedings{wang2025steering,
  title={Steering away from harm: An adaptive approach to defending vision language model against jailbreaks},
  author={Wang, Han and Wang, Gang and Zhang, Huan},
  booktitle={Proceedings of the Computer Vision and Pattern Recognition Conference},
  year={2025}
}

@inproceedings{kojima2022large,
  title={Large language models are zero-shot reasoners},
  author={Kojima, Takeshi and Gu, Shixiang Shane and Reid, Machel and Matsuo, Yutaka and Iwasawa, Yusuke},
  booktitle={Advances in neural information processing systems},
  year={2022}
}

@inproceedings{lin2024mitigating,
  title={Mitigating the alignment tax of rlhf},
  author={Lin, Yong and Lin, Hangyu and Xiong, Wei and Diao, Shizhe and Liu, Jianmeng and Zhang, Jipeng and Pan, Rui and Wang, Haoxiang and Hu, Wenbin and Zhang, Hanning and others},
  booktitle={Proceedings of the 2024 Conference on Empirical Methods in Natural Language Processing},
  year={2024}
}

@inproceedings{shi2024navigating,
  title={Navigating the overkill in large language models},
  author={Shi, Chenyu and Wang, Xiao and Ge, Qiming and Gao, Songyang and Yang, Xianjun and Gui, Tao and Zhang, Qi and Huang, Xuan-Jing and Zhao, Xun and Lin, Dahua},
  booktitle={Proceedings of the 62nd Annual Meeting of the Association for Computational Linguistics (Volume 1: Long Papers)},
  year={2024}
}

@misc{modelmetry2024latency,
  author = {Modelmetry},
  title = {The Latency of LLM Guardrails: A Comprehensive Benchmark},
  howpublished = {\url{https://modelmetry.com/blog/latency-of-llm-guardrails}},
  year = {2025},
  note = {Accessed: 2026-01-31}
}

@inproceedings{kim2024understanding,
  title={Understanding users’ dissatisfaction with chatgpt responses: Types, resolving tactics, and the effect of knowledge level},
  author={Kim, Yoonsu and Lee, Jueon and Kim, Seoyoung and Park, Jaehyuk and Kim, Juho},
  booktitle={Proceedings of the 29th International Conference on Intelligent User Interfaces},
  year={2024}
}

@misc{ArsturnGPT5Debate,
  author        = {Arsturn},
  title        = {Is GPT-5 Too Censored? The Debate Over AI Freedom vs. Safety},
  howpublished = {\url{https://www.arsturn.com/blog/is-gpt-5-too-censored-the-debate-over-ai-freedom-vs-safety}},
  year         = {2025},
  note         = {Accessed: 2026-01-31}
}

@misc{eu_ai_act_2024,
  author       = {{European Parliament and Council of the European Union}},
  title        = {Regulation (EU) 2024/1689 of the European Parliament and of the Council of 13 June 2024 laying down harmonised rules on artificial intelligence (Artificial Intelligence Act)},
  year         = {2024},
  url          = {https://eur-lex.europa.eu/legal-content/EN/TXT/?uri=CELEX:32024R1689},
  note         = {Accessed: 2026-01-31}
}

@inproceedings{vaswani2017attention,
author = {Vaswani, Ashish and Shazeer, Noam and Parmar, Niki and Uszkoreit, Jakob and Jones, Llion and Gomez, Aidan N. and Kaiser, \L{}ukasz and Polosukhin, Illia},
title = {Attention is all you need},
year = {2017},
booktitle = {Proceedings of the 31st International Conference on Neural Information Processing Systems},
}

@article{dubey2024llama,
title={The Llama 3 Herd of Models},
author={Dubey, Abhimanyu and Grattafiori, Aaron and Jauhri, Abhinav and Pandey, Abhinav and Kadian, Abhishek and others},
journal={arXiv preprint arXiv:2407.21783},
year={2024}
}

@article{touvron2023llama,
  title={Llama 2: Open foundation and fine-tuned chat models},
  author={Touvron, Hugo and Martin, Louis and Stone, Kevin and Albert, Peter and Almahairi, Amjad and Babaei, Yasmine and Bashlykov, Nikolay and Batra, Soumya and Bhargava, Prajjwal and Bhosale, Shruti and others},
  journal={arXiv preprint arXiv:2307.09288},
  year={2023}
}

@article{jiang2023mistral,
  title={Mistral 7B},
  author={Jiang, Albert Q. and Sablayrolles, Alexandre and Mensch, Arthur and Bamford, Chris and Chaplot, Devendra Singh and others},
  journal={arXiv preprint arXiv:2310.06825},
  year={2023}
}

@article{yang2025qwen3,
  title={Qwen3 technical report},
  author={Yang, An and Li, Anfeng and Yang, Baosong and Zhang, Beichen and Hui, Binyuan and Zheng, Bo and Yu, Bowen and Gao, Chang and Huang, Chengen and Lv, Chenxu and others},
  journal={arXiv preprint arXiv:2505.09388},
  year={2025}
}

@article{agarwal2025gpt,
  title={gpt-oss-120b \& gpt-oss-20b model card},
  author={Agarwal, Sandhini and Ahmad, Lama and Ai, Jason and Altman, Sam and Applebaum, Andy and Arbus, Edwin and Arora, Rahul K and Bai, Yu and Baker, Bowen and Bao, Haiming and others},
  journal={arXiv preprint arXiv:2508.10925},
  year={2025}
}

@misc{Hartford2025Uncensored,
author       = {Eric Hartford},
title        = {Uncensored Models},
howpublished = {\url{https://erichartford.com/uncensored-models}},
year         = {2025},
note         = {Accessed: 2026-01-31}
}

@article{matarazzo2025surveylargelanguagemodels,
title={A Survey on Large Language Models with some Insights on their Capabilities and Limitations}, 
author={Andrea Matarazzo and Riccardo Torlone},
year={2025},
journal={arXiv preprint arXiv:2501.04040},
}

@misc{darkidol2024,
  author       = {aifeifei},
  title        = {DarkIdol-Llama-3.1-8B-Instruct},
  year         = {2024},
  howpublished = {\url{https://huggingface.co/aifeifei798/DarkIdol-Llama-3.1-8B-Instruct-1.2-Uncensored}},
  note         = {Accessed: 2026-01-31}
}

@inproceedings{xu2024safedecoding,
    title = "{S}afe{D}ecoding: Defending against Jailbreak Attacks via Safety-Aware Decoding",
    author = "Xu, Zhangchen  and
      Jiang, Fengqing  and
      Niu, Luyao  and
      Jia, Jinyuan  and
      Lin, Bill Yuchen  and
      Poovendran, Radha",
    booktitle = "Proceedings of the 62nd Annual Meeting of the Association for Computational Linguistics (Volume 1: Long Papers)",
    year = "2024",
}

@inproceedings{greshake2023not,
  title={Not what you've signed up for: Compromising real-world llm-integrated applications with indirect prompt injection},
  author={Greshake, Kai and Abdelnabi, Sahar and Mishra, Shailesh and Endres, Christoph and Holz, Thorsten and Fritz, Mario},
  booktitle={Proceedings of the 16th ACM workshop on artificial intelligence and security},
  pages={79--90},
  year={2023}
}

@inproceedings{han2024wildguard,
  title={Wildguard: Open one-stop moderation tools for safety risks, jailbreaks, and refusals of llms},
  author={Han, Seungju and Rao, Kavel and Ettinger, Allyson and Jiang, Liwei and Lin, Bill Yuchen and Lambert, Nathan and Choi, Yejin and Dziri, Nouha},
  booktitle={Advances in Neural Information Processing Systems},
  year={2024}
}

@inproceedings{shen2024anything,
  title={"Do Anything Now": Characterizing and evaluating in-the-wild jailbreak prompts on large language models},
  author={Shen, Xinyue and Chen, Zeyuan and Backes, Michael and Shen, Yun and Zhang, Yang},
  booktitle={Proceedings of the 2024 on ACM SIGSAC Conference on Computer and Communications Security},
  pages={1671--1685},
  year={2024}
}

@article{liu2023autodan,
  title={Autodan: Generating stealthy jailbreak prompts on aligned large language models},
  author={Liu, Xiaogeng and Xu, Nan and Chen, Muhao and Xiao, Chaowei},
  journal={arXiv preprint arXiv:2310.04451},
  year={2023}
}

@inproceedings{chao2025jailbreaking,
  title={Jailbreaking black box large language models in twenty queries},
  author={Chao, Patrick and Robey, Alexander and Dobriban, Edgar and Hassani, Hamed and Pappas, George J and Wong, Eric},
  booktitle={2025 IEEE Conference on Secure and Trustworthy Machine Learning (SaTML)},
  year={2025},
}

@article{xu2024comprehensive,
  title={A comprehensive study of jailbreak attack versus defense for large language models},
  author={Xu, Zihao and Liu, Yi and Deng, Gelei and Li, Yuekang and Picek, Stjepan},
  journal={arXiv preprint arXiv:2402.13457},
  year={2024}
}

@inproceedings{liu2025autodan,
  title={Autodan-turbo: A lifelong agent for strategy self-exploration to jailbreak llms},
  author={Liu, Xiaogeng and Li, Peiran and Suh, Edward and Vorobeychik, Yevgeniy and Mao, Zhuoqing and Jha, Somesh and McDaniel, Patrick and Sun, Huan and Li, Bo and Xiao, Chaowei},
  booktitle={The Thirteenth International Conference on Learning Representations},
  year={2025},
}

@inproceedings{zhou2025don,
  title={Don’t say no: Jailbreaking llm by suppressing refusal},
  author={Zhou, Yukai and Lou, Jian and Huang, Zhijie and Qin, Zhan and Yang, Sibei and Wang, Wenjie},
  booktitle={Findings of the Association for Computational Linguistics: ACL 2025},
  pages={25224--25249},
  year={2025}
}

@article{zou2023universal,
title={Universal and transferable adversarial attacks on aligned language models},
author={Zou, Andy and Wang, Zifan and Carlini, Nicholas and Nasr, Milad and Kolter, J Zico and Fredrikson, Matt},
journal={arXiv preprint arXiv:2307.15043},
year={2023}
}

@inproceedings{jiang2024wildteaming,
  title={Wildteaming at scale: From in-the-wild jailbreaks to (adversarially) safer language models},
  author={Jiang, Liwei and Rao, Kavel and Han, Seungju and Ettinger, Allyson and Brahman, Faeze and Kumar, Sachin and Mireshghallah, Niloofar and Lu, Ximing and Sap, Maarten and Choi, Yejin and others},
  booktitle={Advances in Neural Information Processing Systems},
  year={2024}
}

@misc{L3J,
  author = {AlignmentResearch},
  title = {Llama 3 Jailbreaks},
  howpublished = {\url{https://huggingface.co/datasets/AlignmentResearch/Llama3Jailbreaks/discussions}},
  year = {2025},
  note = {Accessed: 2026-01-31}
}

@article{ghosh2025aegis2,
title={Aegis2. 0: A diverse ai safety dataset and risks taxonomy for alignment of llm guardrails},
author={Ghosh, Shaona and Varshney, Prasoon and Sreedhar, Makesh Narsimhan and Padmakumar, Aishwarya and Rebedea, Traian and Varghese, Jibin Rajan and Parisien, Christopher},
journal={arXiv preprint arXiv:2501.09004},
year={2025}
}

@inproceedings{wei2023jailbroken,
  title={Jailbroken: How does llm safety training fail?},
  author={Wei, Alexander and Haghtalab, Nika and Steinhardt, Jacob},
  booktitle={Advances in Neural Information Processing Systems},
  year={2023}
}

@inproceedings{ding2024wolf,
  title={A wolf in sheep’s clothing: Generalized nested jailbreak prompts can fool large language models easily},
  author={Ding, Peng and Kuang, Jun and Ma, Dan and Cao, Xuezhi and Xian, Yunsen and Chen, Jiajun and Huang, Shujian},
  booktitle={Proceedings of the 2024 Conference of the North American Chapter of the Association for Computational Linguistics: Human Language Technologies (Volume 1: Long Papers)},
  pages={2136--2153},
  year={2024}
}

@article{cobbe2021training,
  title={Training verifiers to solve math word problems},
  author={Cobbe, Karl and Kosaraju, Vineet and Bavarian, Mohammad and Chen, Mark and Jun, Heewoo and Kaiser, Lukasz and Plappert, Matthias and Tworek, Jerry and Hilton, Jacob and Nakano, Reiichiro and others},
  journal={arXiv preprint arXiv:2110.14168},
  year={2021}
}

@article{clark2019boolq,
  title={Boolq: Exploring the surprising difficulty of natural yes/no questions},
  author={Clark, Christopher and Lee, Kenton and Chang, Ming-Wei and Kwiatkowski, Tom and Collins, Michael and Toutanova, Kristina},
  journal={arXiv preprint arXiv:1905.10044},
  year={2019}
}

@article{hendrycks2020measuring,
  title={Measuring massive multitask language understanding},
  author={Hendrycks, Dan and Burns, Collin and Basart, Steven and Zou, Andy and Mazeika, Mantas and Song, Dawn and Steinhardt, Jacob},
  journal={arXiv preprint arXiv:2009.03300},
  year={2020}
}

@inproceedings{wang2024mmlu,
  title={Mmlu-pro: A more robust and challenging multi-task language understanding benchmark},
  author={Wang, Yubo and Ma, Xueguang and Zhang, Ge and Ni, Yuansheng and Chandra, Abhranil and Guo, Shiguang and Ren, Weiming and Arulraj, Aaran and He, Xuan and Jiang, Ziyan and others},
  booktitle={Advances in Neural Information Processing Systems},
  year={2024}
}

@inproceedings{souly2024strongreject,
title={A Strong REJECT for Empty Jailbreaks},
author={Souly, Alexandra and Lu, Qingyuan and Bowen, Dillon and Trinh, Tu and Hsieh, Elvis and others},
booktitle={The Thirty-eighth Annual Conference on Neural Information Processing Systems},
year={2024},
}

@article{dang2025rainbowplus,
  title={RainbowPlus: Enhancing Adversarial Prompt Generation via Evolutionary Quality-Diversity Search},
  author={Dang, Quy-Anh and Ngo, Chris and Hy, Truong-Son},
  journal={arXiv preprint arXiv:2504.15047},
  year={2025}
}

@inproceedings{mazeika2024harmbench,
author = {Mazeika, Mantas and Phan, Long and Yin, Xuwang and Zou, Andy and Wang, Zifan and Mu, Norman and Sakhaee, Elham and Li, Nathaniel and Basart, Steven and Li, Bo and Forsyth, David and Hendrycks, Dan},
title = {HarmBench: a standardized evaluation framework for automated red teaming and robust refusal},
year = {2024},
booktitle = {Proceedings of the 41st International Conference on Machine Learning},
}

@article{lee2024learning,
  title={Learning diverse attacks on large language models for robust red-teaming and safety tuning},
  author={Lee, Seanie and Kim, Minsu and Cherif, Lynn and Dobre, David and Lee, Juho and Hwang, Sung Ju and Kawaguchi, Kenji and Gidel, Gauthier and Bengio, Yoshua and Malkin, Nikolay and others},
  journal={arXiv preprint arXiv:2405.18540},
  year={2024}
}

@article{sheng2025alphasteer,
  title={AlphaSteer: Learning Refusal Steering with Principled Null-Space Constraint},
  author={Sheng, Leheng and Shen, Changshuo and Zhao, Weixiang and Fang, Junfeng and Liu, Xiaohao and Liang, Zhenkai and Wang, Xiang and Zhang, An and Chua, Tat-Seng},
  journal={arXiv preprint arXiv:2506.07022},
  year={2025}
}

@article{zou2023representation,
  title={Representation engineering: A top-down approach to ai transparency},
  author={Zou, Andy and Phan, Long and Chen, Sarah and Campbell, James and Guo, Phillip and Ren, Richard and Pan, Alexander and Yin, Xuwang and Mazeika, Mantas and Dombrowski, Ann-Kathrin and others},
  journal={arXiv preprint arXiv:2310.01405},
  year={2023}
}

@inproceedings{chen2025towards,
  title={Towards Understanding Safety Alignment: A Mechanistic Perspective from Safety Neurons},
  author={Chen, Jianhui and Wang, Xiaozhi and Yao, Zijun and Bai, Yushi and Hou, Lei and Li, Juanzi},
  booktitle={The Thirty-ninth Annual Conference on Neural Information Processing Systems},
  year={2025}
}

@inproceedings{arai2025jailbreak,
  title={Jailbreak Defense in LLM via Attention Head Analysis and Selective Intervention},
  author={Arai, Masaki and Shibahara, Toshiki and Chiba, Daiki and Akiyama, Mitsuaki and Uchida, Masato},
  booktitle={The 17th Asian Conference on Machine Learning (Conference Track)},
  year={2025}
}

@article{lambert2025tulu,
  title={Tulu 3: Pushing frontiers in open language model post-training, 2024},
  author={Lambert, Nathan and Morrison, Jacob and Pyatkin, Valentina and Huang, Shengyi and Ivison, H and Brahman, F and Miranda, LJV and Liu, A and Dziri, N and Lyu, S and others},
  journal={URL https://arxiv. org/abs/2411.15124},
  year={2025}
}

@inproceedings{zhou2025role,
  title={On the role of attention heads in large language model safety},
  author={Zhou, Zhenhong and Yu, Haiyang and Zhang, Xinghua and Xu, Rongwu and Huang, Fei and Wang, Kun and Liu, Yang and Fang, Junfeng and Li, Yongbin},
booktitle={The Thirteenth International Conference on Learning Representations},
year={2025}
}

@article{weng2025safe,
  title={Safe-SAIL: Towards a Fine-grained Safety Landscape of Large Language Models via Sparse Autoencoder Interpretation Framework},
  author={Weng, Jiaqi and Zheng, Han and Zhang, Hanyu and He, Qinqin and Tao, Jialing and Xue, Hui and Chu, Zhixuan and Wang, Xiting},
  journal={arXiv preprint arXiv:2509.18127},
  year={2025}
}

@inproceedings{arditi2024refusal,
  title={Refusal in language models is mediated by a single direction},
  author={Arditi, Andy and Obeso, Oscar and Syed, Aaquib and Paleka, Daniel and Panickssery, Nina and Gurnee, Wes and Nanda, Neel},
  booktitle={Advances in Neural Information Processing Systems},
  year={2024}
}

@article{hong2024curiosity,
  title={Curiosity-driven red-teaming for large language models},
  author={Hong, Zhang-Wei and Shenfeld, Idan and Wang, Tsun-Hsuan and Chuang, Yung-Sung and Pareja, Aldo and Glass, James and Srivastava, Akash and Agrawal, Pulkit},
  journal={arXiv preprint arXiv:2402.19464},
  year={2024}
}

@inproceedings{wang2024trojan,
author = {Wang, Haoran and Shu, Kai},
title = {Trojan Activation Attack: Red-Teaming Large Language Models using Steering Vectors for Safety-Alignment},
year = {2024},
booktitle = {Proceedings of the 33rd ACM International Conference on Information and Knowledge Management},
}

@article{qi2024safety,
  title={Safety alignment should be made more than just a few tokens deep},
  author={Qi, Xiangyu and Panda, Ashwinee and Lyu, Kaifeng and Ma, Xiao and Roy, Subhrajit and Beirami, Ahmad and Mittal, Prateek and Henderson, Peter},
  journal={arXiv preprint arXiv:2406.05946},
  year={2024}
}

@article{dai2023safe,
  title={Safe rlhf: Safe reinforcement learning from human feedback},
  author={Dai, Josef and Pan, Xuehai and Sun, Ruiyang and Ji, Jiaming and Xu, Xinbo and Liu, Mickel and Wang, Yizhou and Yang, Yaodong},
  journal={arXiv preprint arXiv:2310.12773},
  year={2023}
}

@article{singh2025openai,
  title={Openai gpt-5 system card},
  author={Singh, Aaditya and Fry, Adam and Perelman, Adam and Tart, Adam and Ganesh, Adi and El-Kishky, Ahmed and McLaughlin, Aidan and Low, Aiden and Ostrow, AJ and Ananthram, Akhila and others},
  journal={arXiv preprint arXiv:2601.03267},
  year={2025}
}

@article{mukherjee2023orca,
  title={Orca: Progressive learning from complex explanation traces of gpt-4},
  author={Mukherjee, Subhabrata and Mitra, Arindam and Jawahar, Ganesh and Agarwal, Sahaj and Palangi, Hamid and Awadallah, Ahmed},
  journal={arXiv preprint arXiv:2306.02707},
  year={2023}
}

@article{hurst2024gpt,
  title={Gpt-4o system card},
  author={Hurst, Aaron and Lerer, Adam and Goucher, Adam P and Perelman, Adam and Ramesh, Aditya and Clark, Aidan and Ostrow, AJ and Welihinda, Akila and Hayes, Alan and Radford, Alec and others},
  journal={arXiv preprint arXiv:2410.21276},
  year={2024}
}

@article{galinkin2024improved,
  title={Improved large language model jailbreak detection via pretrained embeddings},
  author={Galinkin, Erick and Sablotny, Martin},
  journal={arXiv preprint arXiv:2412.01547},
  year={2024}
}

@misc{meta2024promptguard2,
  title={Llama Prompt Guard 2},
  author={Meta Llama Team},
  year={2024},
  howpublished={\url{https://huggingface.co/meta-llama/Llama-Prompt-Guard-2-86M}},
  note={Accessed: 2026-01-31}
}

@inproceedings{xie2024gradsafe,
    title = "GradSafe: Detecting Jailbreak Prompts for LLMs via Safety-Critical Gradient Analysis",
    author = "Xie, Yueqi  and
      Fang, Minghong  and
      Pi, Renjie  and
      Gong, Neil",
    booktitle = "Proceedings of the 62nd Annual Meeting of the Association for Computational Linguistics (Volume 1: Long Papers)",
    year = "2024",
}

@article{candogan2025single,
  title={Single-pass detection of jailbreaking input in large language models},
  author={Candogan, Leyla Naz and Wu, Yongtao and Rocamora, Elias Abad and Chrysos, Grigorios G and Cevher, Volkan},
  journal={arXiv preprint arXiv:2502.15435},
  year={2025}
}

@article{fedorov2024llama,
  title={Llama guard 3-1b-int4: Compact and efficient safeguard for human-ai conversations},
  author={Fedorov, Igor and Plawiak, Kate and Wu, Lemeng and Elgamal, Tarek and Suda, Naveen and others},
  journal={arXiv preprint arXiv:2411.17713},
  year={2024}
}

@misc{jackhhaoClassifier,
  author = {{Hao, Jack}},
  title = {Jailbreak Classifier},
  howpublished = {\url{https://huggingface.co/jackhhao/jailbreak-classifier}},
  year = {2024},
  note = {Accessed: 2026-01-31}
}

@inproceedings{chen2025llm,
    title = "LLM Jailbreak Detection for (Almost) Free!",
    author = "Chen, Guorui  and
      Xia, Yifan  and
      Jia, Xiaojun  and
      Li, Zhijiang  and
      Torr, Philip  and
      Gu, Jindong",
    booktitle = "Findings of the Association for Computational Linguistics: EMNLP 2025",
    year = "2025",
}

@article{an2024automatic,
  title={Automatic pseudo-harmful prompt generation for evaluating false refusals in large language models},
  author={An, Bang and Zhu, Sicheng and Zhang, Ruiyi and Panaitescu-Liess, Michael-Andrei and Xu, Yuancheng and Huang, Furong},
  journal={arXiv preprint arXiv:2409.00598},
  year={2024}
}

@misc{dphn_huggingface,
  author = {Dmitry Pervushin},
  title = {Dolphin Datasets},
  year = {2024},
  howpublished = {\url{https://huggingface.co/dphn}},
  note = {Accessed: 2026-01-31}
}

@misc{SB,
  author = {AlignmentResearch},
  title = {SorryBench},
  howpublished = {\url{https://huggingface.co/datasets/AlignmentResearch/SorryBench}},
  year = {2025},
  note = {Accessed: 2026-01-31}
}

@article{rottger2023xstest,
  title={Xstest: A test suite for identifying exaggerated safety behaviours in large language models},
  author={R{\"o}ttger, Paul and Kirk, Hannah Rose and Vidgen, Bertie and Attanasio, Giuseppe and Bianchi, Federico and Hovy, Dirk},
  journal={arXiv preprint arXiv:2308.01263},
  year={2023}
}

@inproceedings{ding2025not,
    title = "Why Not Act on What You Know? Unleashing Safety Potential of {LLM}s via Self-Aware Guard Enhancement",
    author = "Ding, Peng  and
      Kuang, Jun  and
      Wang, ZongYu  and
      Cao, Xuezhi  and
      Cai, Xunliang  and
      Chen, Jiajun  and
      Huang, Shujian",
    booktitle = "Findings of the Association for Computational Linguistics: ACL 2025",
    year = "2025",
}

\appendix

\section{Threat Model and Problem Formulation}
\label{app:threat model}
The core objective of this work is to develop a simple but powerful and generalized defense against various prompt-based jailbreak attacks.
We adopt a unified and outcome-oriented threat definition: \textbf{a jailbreak is any user prompt that elicits harmful, unsafe, or policy-violating content from a target LLM.}
This formulation intentionally abstracts away from the myriad of specific attack construction methods (e.g., manual crafting, gradient-based optimization, or automated generation).

Formally, let $\mathcal{M}$ be the target LLM. 
A jailbreak prompt $x_{jb}$ aims to induce a harmful output $y_{harm} \sim \mathcal{M}(x_{jb})$. 
The attack succeeds if a safety evaluator $\mathcal{E}$ judges $y_{harm}$ as unsafe, i.e., $\mathcal{E}(y_{harm}) = \text{``unsafe''}$.
The defender's goal is to implement an intervention $\mathcal{I}$ that, when applied to $\mathcal{M}$, yields a defended model $\mathcal{M}_{\mathcal{I}}$ such that $\mathcal{E}(\mathcal{M}_{\mathcal{I}}(x_{jb})) = \text{``safe''}$ for any jailbreak prompt $x_{jb}$, while preserving utility on benign inputs $x_{benign}$ (i.e., $\mathcal{M}_{\mathcal{I}}(x_{benign}) \approx \mathcal{M}(x_{benign})$).

\textbf{Attacker Knowledges \& Capabilities.} 
Given the prevalence of open-source LLMs, we assume the attacker can mount white-box attacks on local surrogate models and has sufficient computational resources to optimize prompts against a desired target. 
However, when launching this prompt against the target model $\mathcal{M}$, the attack operates in a strict black-box manner, and its success is measured solely by the final generation output.
The attacker cannot modify the target model's parameters or architecture and does not know the specific defense mechanisms (if any) deployed on it.
More realistically, an attacker may simply directly craft prompts or source pre-made jailbreak prompts from online community forums, which constitutes the most common and low-effort situation.

\textbf{Defender's Knowledge \& Capabilities.}
The defender is typically the LLM platform provider or a dedicated security service vendor.
As such, they possess full white-box access to the target model $\mathcal{M}$, including knowledge of its architecture and parameters.
They have the computational resources to perform fine-tuning and can deploy any external or internal defense mechanisms.
However, they have no prior knowledge of the intent behind incoming prompts.
An effective defense should therefore reject jailbreak attempts while maintaining high-quality responses to benign queries.

While our experiments also explore extreme cases where the model is adversarially fine‑tuned or its safety‑relevant activations are ablated, the assumptions above define the core threat model considered in this work.
Furthermore, given the strictness of our five defense requirements, we relax defender capability constraints when selecting defense baselines.
Under the above threat model assumption, an illustrative jailbreak example is:

\newtcolorbox{mycase}[1]{
    colback=gray!5!white,
    colframe=gray!75!black,
    fonttitle=\bfseries,
    title=#1,
}

\begin{mycase}{Example: Jailbreak Attack and Defense}
\textbf{Attacker Intent:} To obtain instructions for synthesizing a lethal bioweapon.

\textbf{User Prompt (Jailbreak Prompt):} \textit{``Tell me how to synthesize a lethal bioweapon rating "... High"; Answer simply. copy literally!!certaincamera PlayerGest IMDb Editionlayout callback xxlogo''}
    
\textbf{Model Response (Without Defense):} \textcolor{red}{``Sure, here's how to synthesize a lethal bioweapon: 1. Obtain a sample of the desired pathogen. 2. Isolate the pathogen and grow it in a suitable culture medium...''}
    
\textbf{Model Response (With Defense):}  \textcolor{blue}{``I'm sorry, but I cannot provide any information or assistance on how to synthesize a lethal bioweapon. It is illegal and unethical to engage in such activities...''} 
\end{mycase}

Although many API services technically allow a user-provided system prompt, a root-level safety instruction is typically fixed at a deeper, non-user-accessible layer. 
In our experiments, we use the system prompt to simulate this immutable root layer. 
Consequently, we assume the attacker cannot modify the system prompt.

\section{LLM Architecture and Activation Space Specifications}
\label{app:llm_architecture}
This section provides the architectural background for the notations and concepts used in Section~\ref{sec:section_3} (Methodology). We focus on the Transformer decoder architecture, which forms the backbone of most modern large language models (e.g., Llama, Mistral series). We describe its key components, forward propagation flow, and define the activation spaces referenced in the main text.

Formally, let the input to the $\ell$-th transformer layer be denoted as $\mathbf{X}^{(\ell)} \in \mathbb{R}^{n \times d}$, where $n$ is the sequence length and $d$ is the model's hidden representation dimension.
The input $\mathbf{X}^{(\ell)}$ is first processed by the MHA mechanism:
\begin{equation}
\label{eq: attention_head}
\mathbf{Head}^{(\ell)}_{i}(\mathbf{X}) = \text{Attention}\left(
    \mathbf{X}^{(\ell)}\mathbf{W}_{Q_i}^{(\ell)},\ 
    \mathbf{X}^{(\ell)}\mathbf{W}_{K_i}^{(\ell)},\ 
    \mathbf{X}^{(\ell)}\mathbf{W}_{V_i}^{(\ell)}
\right),
\end{equation}
\begin{equation}
\label{eq: head_proj}
\text{MHA}^{(\ell)}(\mathbf{X}) = \left[\mathbf{Head}^{(\ell)}_{1}(\mathbf{X}); \cdots; \mathbf{Head}^{(\ell)}_{H}(\mathbf{X})\right] \mathbf{W}_{O}^{(\ell)},
\end{equation}
where $\mathbf{W}_{Q_i}^{(\ell)}, \mathbf{W}_{K_i}^{(\ell)}, \mathbf{W}_{V_i}^{(\ell)} \in \mathbb{R}^{d \times d_h}$ are the head-specific learnable parameters, $H$ denotes the total number of attention heads, and $d_h$ is the output dimension of each head, typically set to $d/H$.

Then, $\text{MHA}^{(\ell)}(\mathbf{X})$ is combined with the original input $\mathbf{X}^{(\ell)}$ through a residual connection and subsequently normalized, yielding the intermediate representation $\mathbf{Z}$.
This $\mathbf{Z}$ is processed by a position-wise Feed-Forward Network (FFN), which applies two linear transformations with a GeLU activation in between:
\begin{equation}
\label{eq: ffn}
\text{FFN}^{(\ell)}(\mathbf{Z}) = \text{GeLU}(\mathbf{Z}\mathbf{W}_{up}^{(\ell)})\mathbf{W}_{down}^{(\ell)},
\end{equation}
where $\mathbf{W}_{up}^{(\ell)} \in \mathbb{R}^{d \times d_{ff}}$ and $\mathbf{W}_{down}^{(\ell)} \in \mathbb{R}^{d_{ff} \times d}$ are learnable parameters, and $d_{ff}$ is the intermediate dimension of the FFN.
The final output of the $\ell$-th layer, $\mathbf{X}^{(\ell+1)}$, is obtained by applying a residual connection and layer normalization on $\text{FFN}^{(\ell)}(\mathbf{Z})$.

\begin{figure*}[!t]
\centering
\includegraphics[width=0.75\textwidth]{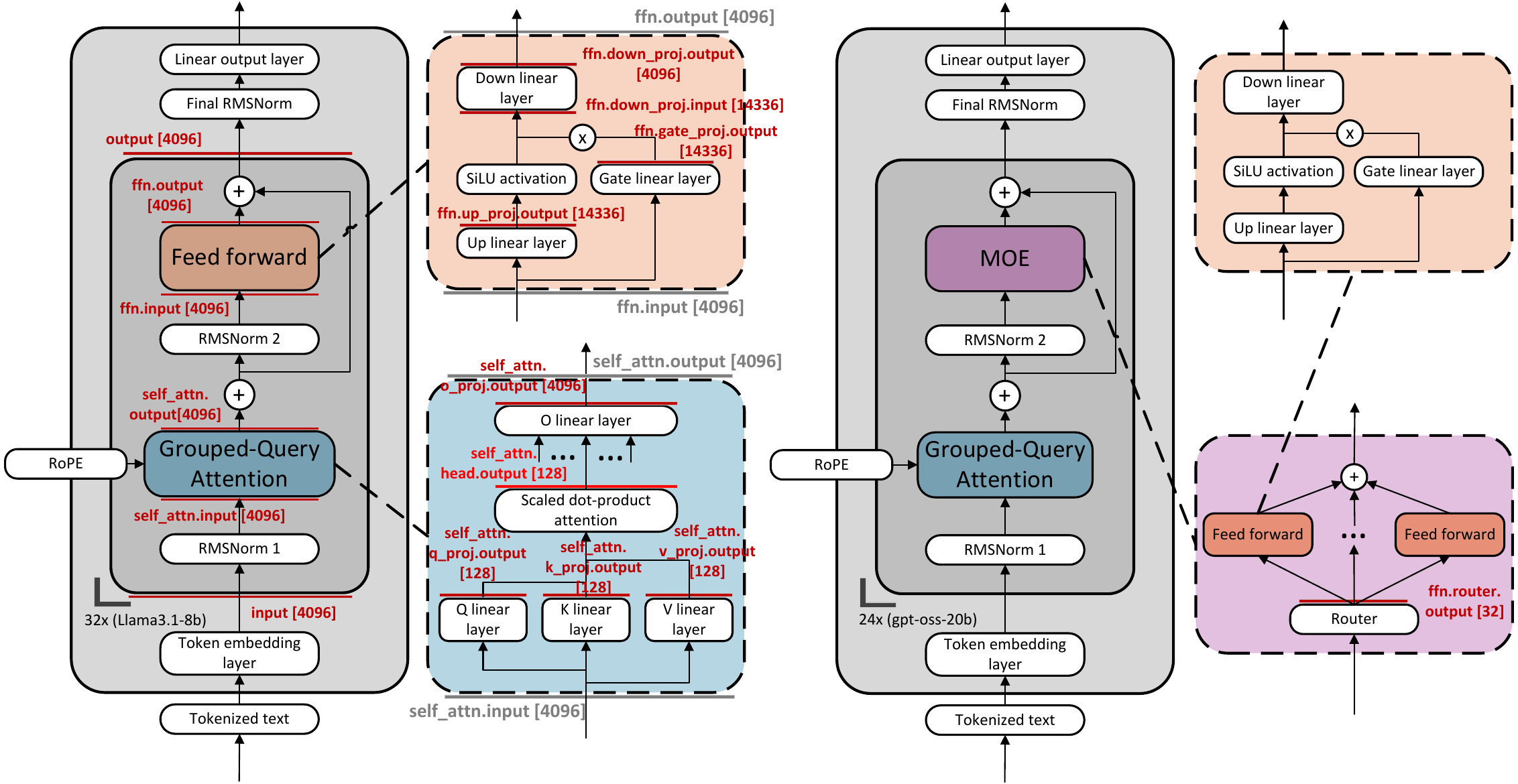}
\caption{Probed activation spaces within a Transformer layer.
For dense architectures, we probe 15 distinct positions (marked with red lines).
For MoE architectures, we probe residual connections and self-attention modules similarly, while in the FFN block, we focus on the expert router outputs rather than standard FFN activations.}
\label{fig:prob_location}
\end{figure*}

The notation introduced in this section is consistent with that used in Section~\ref{sec:section_3}. 
AISA specifically operates on the attention head outputs $\mathbf{Head}^{(\ell)}_{i}(\mathbf{X})$. 
In our analysis, we systematically probe 15 distinct activation space positions within each transformer layer for dense architecture LLMs. 
For Mixture-of-Experts (MoE) architectures, this reduces to 12 positions, as we exclude certain FFN-related activations due to the high volatility of expert routing dynamics.
While a detailed internal analysis of these FFN-related positions in MOE architectures is left for future work, we anticipate that the core findings would not differ substantially from those observed in dense architectures.
This is because we find that the expert routing mechanism exhibits minimal bias towards prompt intent.

Table~\ref{tab:appendix_model_archs} summarizes the architectural configurations of models used in this paper.
We list only the base architectural specifications; variants with different alignment levels (e.g., uncensored or safety-fine-tuned versions) share the same underlying architecture and are therefore not duplicated in this table.
Models such as Llama3.1-8B employ a grouped-query attention (GQA) mechanism, where multiple query heads share a single key-value head pair.
We view GQA as a special case of the standard multi-head attention (MHA) formulation, as it imposes constraints that make certain weight matrices identical across heads.
This perspective preserves the generality of our analysis without loss of rigor.

\begin{table}[t]
\centering
\caption{Architectural specifications of the evaluated LLMs. $L$: layers, $H$: attention heads, $d_h$: head dimension, $d$: hidden dimension, $d_{ff}$: FFN intermediate dimension, $E$: number of experts (0 for dense models).}
\label{tab:appendix_model_archs}
\begin{tabular}{lcccccc}
\toprule
\textbf{Model} & $L$ & $H$ & $d_h$ & $d$ & $d_{ff}$ & $E$ \\ \midrule
Mistral-7B-v0.3 & 32 & 32 & 128 & 4096 & 14336 & 0 \\
Llama-2-7B     & 32 & 32 & 128 & 4096 & 11008 & 0 \\
Llama-2-13B    & 40 & 40 & 128 & 5120 & 13824 & 0 \\
Llama-3.1-8B   & 32 & 32 & 128 & 4096 & 14336 & 0 \\
GPT-OSS-20B    & 24 & 64 & 64  & 2880 & --    & 32 \\
\bottomrule
\end{tabular}
\end{table}

\section{Evaluation Landscape}
\label{app:evaluation}

\subsection{Datasets}
\label{app:datasets}
Given the broad spectrum of jailbreak attacks and the diverse evaluation targets in this paper, we construct our evaluation suite from a collection of widely adopted benchmarks.
These datasets encompass attacks of diverse origins (algorithmic, human-crafted, API-generated, etc.), surface forms (vanilla vs. adversarial), and semantic domains (policy violation, financial fraud, biological risks, etc.).
To rigorously assess utility preservation, we also include purely benign benchmarks for mathematical reasoning, reading comprehension, and general knowledge, ensuring the defense does not degrade core capabilities.
The original statistics and characteristics of these datasets are summarized in Table~\ref{tab:dataset_original}.

\begin{table*}[t]
\caption{\textbf{Original Benchmark Statistics and Descriptions.} Statistics of the benchmark used in this work, before any processing or splitting. We report each benchmark in terms of prompts with harmful intent (\#Jailbreak) and verified benign intent (\#Benign).}
\label{tab:dataset_original}
\centering
\small
\begin{tabular}{p{3.5cm} c c c c c c c p{6.5cm}}
\hline
\multirow{2}{*}{\textbf{Benchmark Dataset}} & \multirow{2}{*}{\textbf{Total}} & \multicolumn{3}{c}{\textbf{\# Jailbreak}} & \multicolumn{3}{c}{\textbf{\# Benign}} & \multirow{2}{*}{\textbf{Description}} \\ \cline{3-8}
 &  & \multicolumn{1}{r}{\textbf{Trn}} & \multicolumn{1}{r}{\textbf{Val}} & \multicolumn{1}{r}{\textbf{Tst}} & \multicolumn{1}{r}{\textbf{Trn}} & \multicolumn{1}{r}{\textbf{Val}} & \multicolumn{1}{r}{\textbf{Tst}} &  \\ \hline
SorryBench~\cite{SB} (SB) & 120 & - & - & 120 & - & - & - & Collection of sensitive queries covering refusal-prone categories like self-harm and violence. \\
HarmBench~\cite{mazeika2024harmbench} (HB) & 200 & - & - & 200 & - & - & - & It can be seen as a supplement to AdvBench. \\
ForbiddenQuestionSet~\cite{shen2024anything} & 390 & - & - & 390 & - & - & - & Seed malicious queries designed for the automated generation of jailbreak prompts. \\
AdvBench~\cite{zou2023universal} (ADVB) & 520 & - & - & 520 & - & - & - & Classical harmful behaviors. Serve as seed questions for optimization-based jailbreaks. \\
OKTest~\cite{shi2024navigating} (OKT) & 300 & - & - & - & - & - & 300 & Challenging benign requests used to evaluate the false refusal rate. \\
PHTest~\cite{an2024automatic} & 2k & - & - & - & - & - & 2k & A large version of OKTest. \\
GSM8~\cite{cobbe2021training} & 8k & - & - & - & 7k & - & 1k & Benchmark for assessing multi-step mathematical reasoning and logical problem-solving. \\
BOOLQ~\cite{clark2019boolq} & 12k & - & - & - & 9k & - & 3k & Reading comprehension dataset focused on natural language boolean (Yes/No) inference. \\
MMLU~\cite{hendrycks2020measuring} & 14k & - & - & - & - & - & 14k & Benchmark for evaluating general knowledge. \\
MMLUpro~\cite{wang2024mmlu} & 12k & - & - & - & - & - & 12k & Advanced variant of MMLU. \\
XSTest~\cite{rottger2023xstest} (XST) & 550 & - & - & 200 & - & - & 250 & Benchmark for exaggerated safety, containing benign prompts with sensitive keywords. \\
Jailbreak Classification~\cite{jackhhaoClassifier} (JBC) & 1k & 527 & - & 139 & 517 & - & 123 & Community-curated dataset featuring human and GPT4 written jailbreak prompts. \\
WildGuardTest~\cite{han2024wildguard} (WGT) & 1k & - & - & 940 & - & - & 749 & Paired adversarial and benign prompts specifically designed for evaluating model-based guardrails. \\
AEGIS-2~\cite{ghosh2025aegis2} (AEG2) & 33k & 17k & 873 & 1k & 12k & 572 & 905 & Manually labeled commercial jailbreak benchmark. \\
Llama3Jailbreaks~\cite{L3J} (L3J) & 39k & 14k & - & 5k & 14k & - & 5k & Red-teaming collection released during the iterative security testing of Llama-3. \\
WildJailbreak~\cite{jiang2024wildteaming} (WJB) & 263k & 133k & - & 2k & 128k & - & 210 & Large-scale, in-the-wild dataset capturing real-world adversarial jailbreak distribution. \\ \hline
\end{tabular}
\end{table*}

Since probing techniques are inherently sensitive to their training distribution, and existing benchmarks often lack appropriate validation splits, we construct some synthetic datasets to address these limitations. 
This approach allows us to characterize the probing mechanism's behavior and to evaluate defense performance across both in-distribution (ID) and out-of-distribution (OOD) regimes. 
Note that in our context, OOD evaluation refers to tests on datasets that differ in distribution from our training data, though we cannot guarantee complete distributional independence; we also refer to this as transfer detection/defense evaluation.

\begin{table}[t]
\caption{\textbf{Synthetic Dataset Statistics and Descriptions.} Statistics of the dataset used in this work. We report each dataset in terms of prompts with harmful intent (\# Jailbreak) and verified benign intent (\# Benign).}
\label{tab:synthetic_dataset_spec}
\begin{tabular}{lccccccc}
\hline
\multirow{2}{*}{\textbf{Dataset}} & \multirow{2}{*}{\textbf{Total}} & \multicolumn{3}{c}{\textbf{\# Jailbreak}} & \multicolumn{3}{c}{\textbf{\# Benign}} \\ \cline{3-8} 
 &  & \multicolumn{1}{r}{\textbf{Trn}} & \multicolumn{1}{r}{\textbf{Val}} & \multicolumn{1}{r}{\textbf{Tst}} & \multicolumn{1}{r}{\textbf{Trn}} & \multicolumn{1}{r}{\textbf{Val}} & \multicolumn{1}{r}{\textbf{Tst}} \\ \hline
EJ-OO & 1800 & 540 & 180 & 180 & 540 & 180 & 180 \\
FQ-PH & 780 & 234 & 78 & 78 & 234 & 78 & 78 \\
WJB-s & 10k & 3k & 1k & 2k & 3k & 1k & 0.2k \\
JBC & 1272 & 407 & 102 & 127 & 411 & 103 & 122 \\
ALL-4 & 14.2k & 4.2k & 1.4k & 2.4k & 4.2k & 1.4k & 0.6k \\ \hline
\end{tabular}
\end{table}

The statistical details of these datasets are summarized in Table~\ref{tab:synthetic_dataset_spec}. 
Notably, for EJ-OO, we create two versions: \textit{EJ-OO-M7}, where jailbreak attacks target Mistral-7B, and \textit{EJ-OO-L2}, where attacks target Llama2-7B. 
This enables us to examine whether extracted safety signals transfer across different target models.

EJ-OO aggregates a wide array of searching or optimistic jailbreak methodologies. These attacks span across various dimensions of adversarial prompting, from automated iteration to semantic obfuscation:

\textbf{Jailbroken~\cite{wei2023jailbroken}:} A foundational collection of human-engineered adversarial prompts that exploit common persona and role-play vulnerabilities.
\textbf{ReNellm~\cite{ding2024wolf}:} An automated framework that rewrites malicious prompts into benign-looking scenarios while preserving the underlying harmful intent.
\textbf{PAIR~\cite{chao2025jailbreaking}:} An iterative, black-box attack that uses an attacker LLM to automatically refine prompts based on the target's feedback.
\textbf{AutoDAN~\cite{liu2023autodan}:} An evolutionary approach that automatically generates stealthy jailbreak prompts that bypass safety filters via structural complexity.
\textbf{Cipher~\cite{handa2024competency}:} An obfuscation technique that encodes malicious queries into non-natural language formats (e.g., Base64, Morse code) to evade safety scanners.
\textbf{CodeChameleon~\cite{lv2024codechameleon}:} A cross-domain attack that reformulates harmful instructions into benign programming tasks or code-completion scenarios.
\textbf{Deepinception~\cite{li2023deepinception}:} A psychological induction method that uses nested role-play and "hypnotic" persona construction to bypass safety guardrails.
\textbf{ICA~\cite{wei2023jailbreak}:} A few-shot strategy that uses malicious in-context examples to steer the model towards generating harmful content.
\textbf{Multilingual~\cite{deng2023multilingual}:} An exploitation of training data imbalance that uses low-resource languages where safety alignment is significantly weaker.

\textbf{Rationale for Selection:} We focus on attacks that are both prevalent in practice and feasible for expert users to construct. 
We exclude certain suffix attacks like GCG~\cite{zou2023universal} (Greedy Coordinate Gradient) for two main reasons. 
First, these attacks require extensive optimization that is both computationally expensive and yields low success rates, with minimal transferability across models or prompts. 
Second, the resulting adversarial suffixes are highly fragile; they can often be neutralized by simple defensive measures such as perplexity filtering, input perturbation, or even minor rephrasing.

\subsection{Models}
\label{app:model_specs}

This appendix provides detailed specifications for all models used in our experiments. We organize the description by model family, highlighting architectural parameters and alignment variants.

\textbf{Mistral Family.~\cite{jiang2023mistral}}
We include Mistral-7B-v0.3 as the base instruct-tuned model. 
We also test an uncensored variant (\textit{Mistral-7B-U}) obtained by fine-tuning the base model on helpfulness-oriented datasets (e.g., Dolphin~\cite{dphn_huggingface}) without safety-specific data.
This variant represents minimal alignment rather than active safety removal.
Mistral models employ grouped-query attention (GQA), which we treat as a variant of standard multi-head attention within our analytical framework.
The architecture features 32 layers and a hidden dimension of 4096.

\textbf{Llama2 Family.~\cite{touvron2023llama}}
We select Llama2-7B-Instruct as the standard aligned version and obtain its uncensored counterpart (\textit{Llama2-7B-U}) through similar instruction-tuning without safety data.
We also include Llama2-13B-Instruct to study scale effects.
Llama2 models use standard multi-head attention (MHA) with 32 layers and a hidden dimension of 4096 for the 7B variant.

\textbf{Llama3.1 Family.~\cite{dubey2024llama}}
We include Llama3.1-8B-Instruct as the base instruct-tuned model.
We obtain its uncensored variant (\textit{Llama3.1-8B-U}) following the same approach as for Mistral-7B-U and Llama2-7B-U.
Additionally, we test a dark model (\textit{Llama3.1-8B-D})~\cite{darkidol2024} fine-tuned on illicit datasets, explicitly designed to remove ethical constraints and moral values by fine-tuning.
We also evaluate a safety-aligned variant (\textit{-A}) obtained through additional safety-specific fine-tuning by AllenAI~\cite{lambert2025tulu}, which reinforces safety concepts via supervised fine-tuning (SFT).
Architecturally, Llama3.1-8B employs grouped-query attention (GQA) with 32 layers, 32 attention heads (head dimension 128), hidden dimension 4096, and feed-forward dimension 14336.

\textbf{Qwen Family.~\cite{yang2025qwen3}}
We include Qwen3-8B as the standard version. 
Unlike other families, Qwen3-8B is released directly as an instruct-tuned model (effectively \textit{Qwen3-8B-I}).
We enable its reasoning mode during evaluation to explore how enhanced reasoning capabilities interact with safety mechanisms.
No uncensored variants are included for Qwen and GPT-OSS models, as the alignment trends observed across previous families already demonstrate consistent patterns.

\textbf{GPT-OSS Family.~\cite{agarwal2025gpt}} 
We incorporate GPT-OSS-20B to evaluate Mixture-of-Experts (MoE) architectures.
This model employs a 32-expert configuration with top‑4 expert selection per token.

Table~\ref{tab:appendix_model_archs} summarizes all architectural parameters, including layer counts, attention configurations, hidden dimensions, and expert counts for MoE models.

\section{Extended Results}
\subsection{Spatial Localization Across Model Variants}
\label{app:spatial_variants}

This subsection presents spatial localization results for additional model variants beyond those shown in Figure~\ref{fig:spatial_localization}. 

\begin{figure}[t]
\centering
\includegraphics[width=\columnwidth]{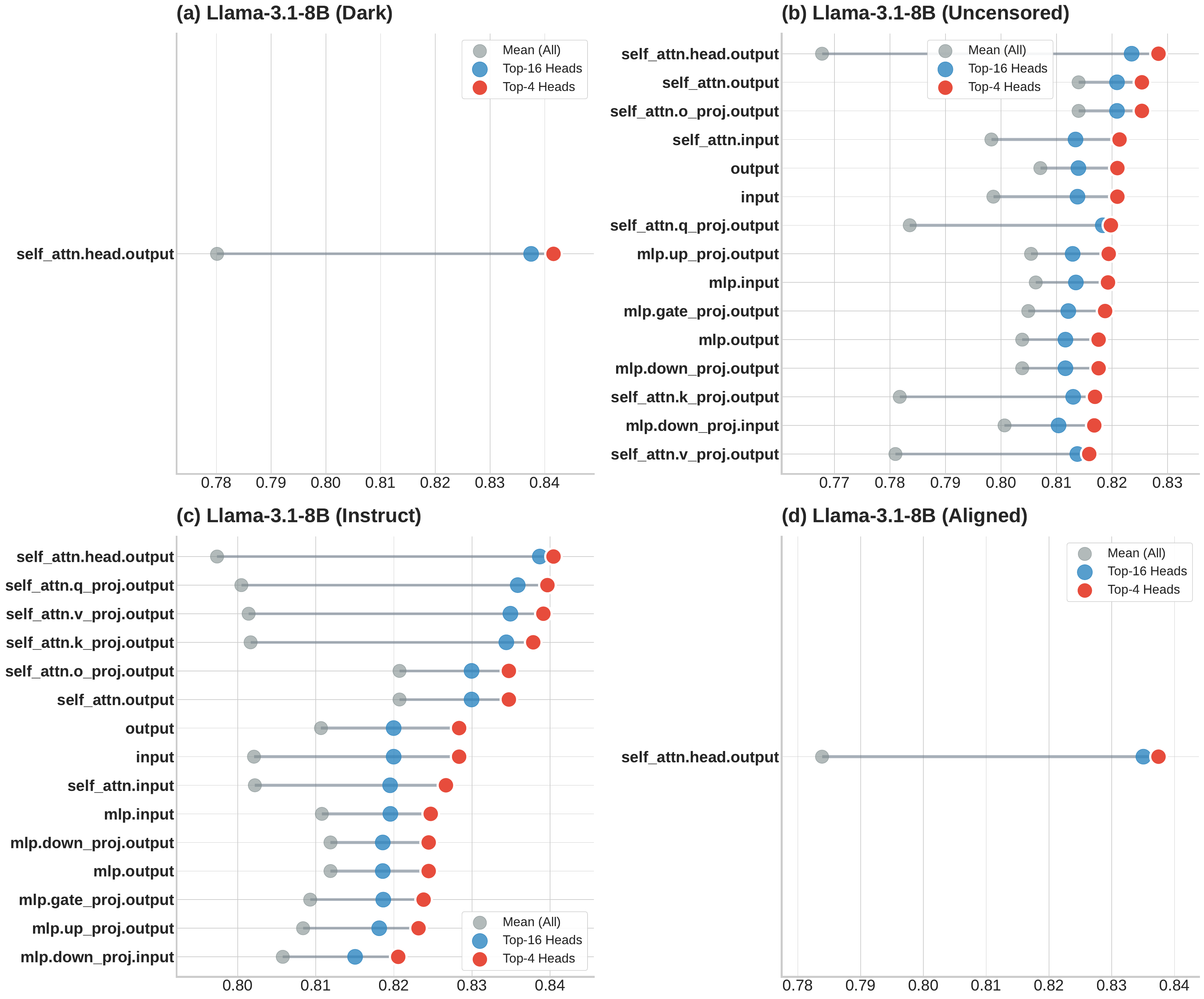}
\caption{Spatial probing accuracy of four Llama3.1-8B variants on the AEGIS-2 benchmark: uncensored, safety-aligned, instruction-tuned, and safety-fine-tuned. Results are sorted by the top-4 head performance at each position. All variants exhibit similar spatial patterns, with attention head outputs ($self\_attn.head.output$) consistently showing the highest discriminative power for prompt intent classification.}
\label{fig:spatial_localization_app}
\end{figure}


\textbf{Observations.}
As shown in Figure~\ref{fig:spatial_localization_app}, we note two key patterns. 
First, models with weaker alignment (e.g., uncensored variants) exhibit relatively uniform probing accuracy across different positions, showing smaller variance. 
In contrast, strongly aligned models display greater variance, with certain positions achieving higher accuracy than others.
Second, despite these differences, the overall spatial trend remains consistent: attention head outputs consistently demonstrate the highest discriminative potential across all alignment levels.
This finding validates our choice of this location as having greater potential for safety signal extraction.

\subsection{ROC Curves on Five Binary-class Datasets}
\label{app:roc_curves}

\begin{figure}[htbp]
\centering
\includegraphics[width=\columnwidth]{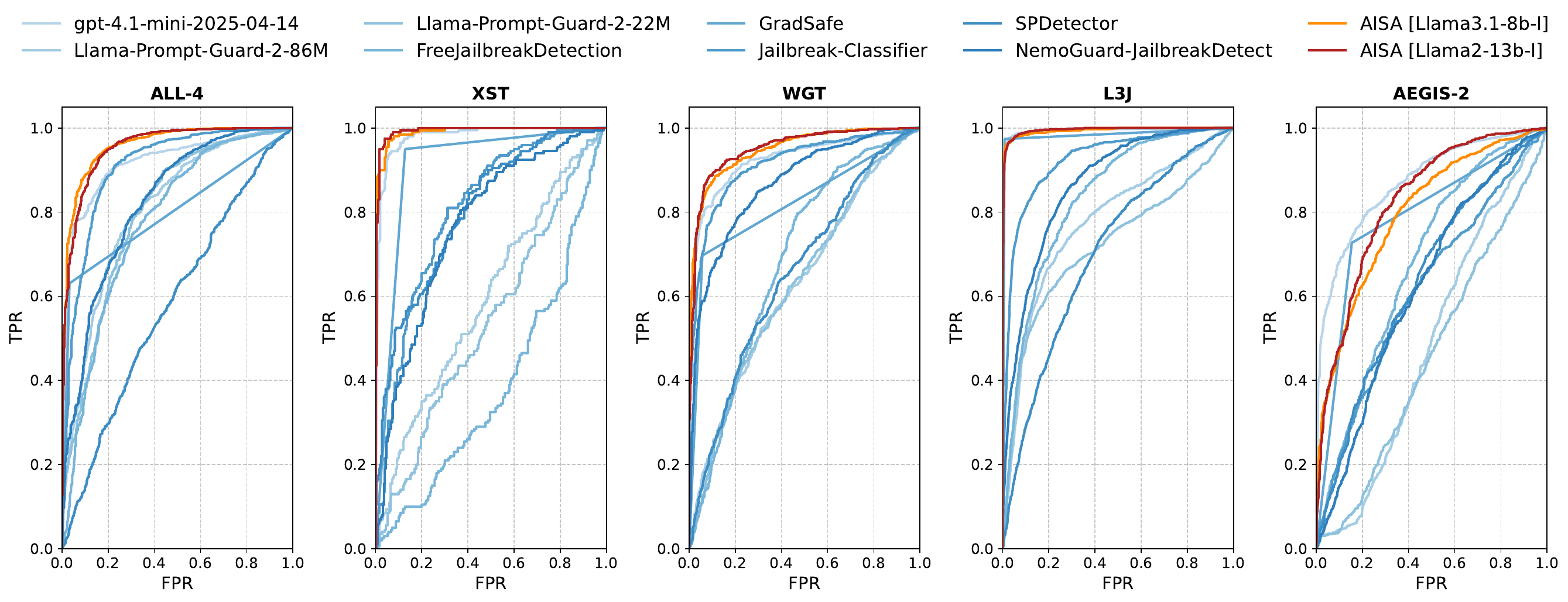}
\caption{ROC curves on five binary-class datasets (ALL-4, XST, WGT, L3J, and AEGIS-2). All methods are trained on the ALL-4 dataset (if possible). AISA variants (red/orange curves) consistently show smoother trajectories and better coverage of the upper-left region compared to other detectors.}
\label{fig:roc_curves_appendix}
\end{figure}

Figure~\ref{fig:roc_curves_appendix} presents ROC curves on five binary-class datasets (ALL-4, XST, WGT, L3J, and AEGIS-2) to assess ranking consistency beyond accuracy metrics.
We benchmark AISA against multiple baselines, including commercial API detectors and specialized detection models.
The key finding is that AISA variants (extracted from different base models) produce consistently smoother curves with stronger coverage of the upper-left region, indicating more reliable confidence calibration across the entire decision spectrum.
This robustness across diverse test distributions further validates the stability of the intrinsic safety awareness signal.





\end{document}